\documentclass[12pt]{article}
\usepackage[T2B]{fontenc}

 \usepackage{amssymb}

\newcommand{\R}{\mathbb{R}}

\newcommand{\CC}{\mathbb{C}}

 \newcommand{\beqn}{\begin{eqnarray}}
 \newcommand{\eeqn}{\end{eqnarray}}
 \newcommand{\be}{\begin{equation}}
 \newcommand{\ee}{\end{equation}}
 \newcommand{\ba}{\begin{array}}
 \newcommand{\ea}{\end{array}}

 \newcommand{\al}{\alpha}
 \newcommand{\ga}{\gamma}
  \newcommand{\la}{\label}
  \newcommand{\pa}{\partial}
  \newcommand{\rot}{{\rm rot}}

 \newcommand{\bv}{{\bf v}}

 \newcommand{\ds}{\displaystyle}

 \newcommand{\br}{|\kern-.25em|\kern-.25em|}
 \newcommand{\brr}{{|\kern-.15em|\kern-.15em|\kern-.15em}\,}

\newcommand{\const}{\mathop{\rm const}\nolimits}

 \newtheorem{theorem}{Theorem}[section]
 \newtheorem{definition}[theorem]{Definition}
 
 \newtheorem{lemma}[theorem]{Lemma}
 
 \newtheorem{remark}[theorem]{Remark}
 \newtheorem{remarks}[theorem]{Remarks}
 \newtheorem{cor}[theorem]{Corollary}

\newcommand{\bo}{{\hfill\loota}}
\newcommand{\loota}{\hbox{\enspace{\vrule height 7pt depth 0pt width
      7pt}}}
\begin{document}

\begin{titlepage}
\hspace{2cm}
 \begin{center}
{\Large\bf Energy-momentum relation for solitary waves
\medskip\\
  of nonlinear Dirac equations}
\vspace{1cm}\\
{\large T.V. Dudnikova
\footnote{Supported partly by the research grant of
RFBR (12-01-00203).}
}\\
{\small\it
Elektrostal Polytechnical Institute\\
Elektrostal, 144000 Russia\\
e-mail:~ tdudnikov@mail.ru}
\end{center}
\vspace{1cm}

\begin{abstract}
Solitary waves of nonlinear Dirac, Maxwell--Dirac and Klein--Gordon--Dirac
equations  are considered.
We prove that the energy-momentum relation for
 solitary waves coincides with the
Einstein  energy-momentum relation for point particles.
\medskip

\medskip

{\it Key words and phases}: Nonlinear Dirac equation, solitary waves,
Einstein energy-momentum relation,
Maxwell--Dirac and Klein--Gordon--Dirac equations
\end{abstract}
\end{titlepage}

\newpage
%%%%%%%%%%%%%%%%%%%%%%%%%%%%%%%%%%%%%%%%%%%%%%%%%%%%%%%%%%%%%%%%%%
\setcounter{equation}{0}
\section{Introduction}
%%%%%%%%%%%%%%%%%%%%%%%%%%%%%%

The paper concerns the old problem of mathematically
describing elementary particles in  field theory.
%%------------------------------------------
Einstein and Grommer \cite{Ein} suggested that particles could be
described as singularities of  solutions to the field equations.
%%---------------------------------------------------------------------
The generalization of this result to interacting systems of particles was given
by Einstein, Infeld and Hoffmann \cite{EIH}.
%%--------------------------------
Rosen \cite{Rn} was the first who proposed
a description of particles for the coupled Klein--Gordon--Maxwell equations,
which are invariant with respect to the Lorentz group.
%% and to the global gauge group $U(1)$:
Namely, the particle at rest is described by
a finite energy solution that has ``Schr\"odinger's'' form
$\varphi(x)e^{-i\omega t}$ (``nonlinear eigenfunctions'' or ``solitary waves'').
The  particle with the nonzero velocity $v$,  $|v|<1$,
is obtained by the corresponding Lorentz (or Poincar\'e) transformation.
%%-----------------------------
Since the work of Rosen some authors have considered the interaction of classical fields,
looking for localized solutions could be used as models of extended particles.

The existence of  solitary waves has been analyzed %numerically
 by many authors for diverse Lagrangian field theories
 \cite{EGS, Lisi, RanV, Rn, Soler, W}, such that nonlinear Dirac fields,
the Maxwell--Dirac (MD) and Klein--Gordon--Dirac (KGD) equations.
 We describe briefly some results.
\medskip

Nonlinear Dirac equations occur in the attempt to construct relativistic
models of extended particles by means of nonlinear Dirac fields.
The review of such models can be found in \cite{Ran}.
The stationary solutions of nonlinear Dirac equation were extensively studied
in the literature used  variational methods \cite{ES} and a dynamical systems approach \cite{CV,Merle,BCDM}.
For details, see the survey papers \cite{ES2002, ELS, Ran}
and the references therein.
\medskip

 The (MD) equations (see, e.g., \cite{BD, Schiff}) describing
 the interaction of an electron with its own electromagnetic field
 have been widely studied by many authors.
The first results on the local existence and uniqueness of solutions
was obtained by Gross \cite{G}, Chadam \cite{Chadam}, Chadam and Glassey \cite{CG}.
The stationary (localized) solutions of the classical
(MD) system were studied numerically by Wakano \cite{W} and Lisi \cite{Lisi}.
Using variational methods, Esteban,  Georgiev and S\'er\'e \cite{EGS}
have proved the existence of stationary solutions
%of the form $\psi(t,x)=e^{-i\omega t}\varphi(x)$,
with $\omega\in(-m,0)$.
These results were extended by Abenda \cite{Ab} for $\omega\in(-m,m)$.

For the (KGD) equations,
the local existence and uniqueness of solutions
  were proved by  Chadam and Glassey \cite{CG74}.
Numerical results on the  stationary states were obtained by
Ranada and Vazquez in \cite{RanV}. The rigorous proof of the existence
for the stationary solutions was given by Esteban {\it et al.} \cite{EGS}.
For some Lorentz invariant complex scalar fields theories,
the particle-like solutions was studied by Rosen \cite{Rg5,Rg6}.
\medskip

Note that it would be  of importance to
develop a particle-like dynamics for moving solitons.
We make a step in this direction
for relativistic-invariant nonlinear Dirac, (MD) and (KGD) equations.
Namely, we  prove that
the energy-momentum relation coincides with that of
a relativistic particle.
\medskip
%%---------------------------------------

Now we outline the main result in the case  of nonlinear Dirac equations.
We consider the  Dirac equations of the form
\be\la{4.1}
i\dot\psi=-i\al\cdot\nabla\psi+m\beta\psi -g(\bar\psi\psi)\beta\psi.
\ee
We use natural units, in which we have rescaled length and time so that
$\hbar=c=1$.
Here unknown function $\psi\equiv\psi(t,x):\R\times\R^3\to\CC^4$
is four-component Dirac spinor field,
%(i.e., $\psi$ is a function on space-time into spin space),
$m>0$,  $\dot\psi=\pa_t\psi$, $x=(x_1,x_2,x_3)$,
$\nabla=(\pa_1,\pa_2,\pa_3)$,
$\partial_k=\pa/(\pa x_k)$, $k=1,2,3$,
%$\gamma^\mu\pa_\mu=\gamma^0\pa_t+\gamma\cdot\nabla$,
$\al=(\al_1,\al_2,\al_3)$.
 $\al_k,\beta$ are the $4\times4$ complex Pauli-Dirac matrices
(in the standard $2\times 2$ blocks representation)
$$
\alpha_k=\left(\ba{cc}0&\sigma_k\\
\sigma_k&0\ea\right)\quad (k=1,2,3),
\quad \beta=\left(\ba{cc}I&0\\
0&-I\ea\right),
$$
where  $I$ denotes the $2\times2$ unit matrix,
and $\sigma_k$ are Pauli matrices defined as
\beqn\nonumber%\la{sigma}
\sigma_1= \left(\ba{ll}
0 & 1\\1 & 0\\ \ea  \right),\quad
\sigma_2=   \left(\ba{ll}
0 & -i\\i & 0\\
\ea \right),\quad
\sigma_3= \left(\ba{ll}1 & 0\\
0 & -1\\\ea   \right).
\eeqn
One verifies that
$\sigma_k\sigma_l+\sigma_l\sigma_k=2\delta_{kl}I$, $\sigma_k^*=\sigma_k$,
$k=1,2,3$.
Then
\be\label{Pmat}
\beta^*=\beta,\quad \al_k^*=\al_k,\quad \al_k^2=\beta^2=I,\quad
\al_k\al_j+\al_j\al_k=0 \quad\mbox{for }\,j\not=k,
\quad \al_k\beta+\beta\al_k=0.
\ee
Let us fix the following notations.
Given two vectors of $\CC^4$, $\psi\phi:=\psi\cdot\phi$
is the inner product in $\CC^4$, $*$ denotes the complex conjugate.
By definition, the "adjoint spinor" is $\bar\psi=\psi^*\beta$.

%%%%%%%%%%%%%%%%%%%%%%%%%%%%%%
The particular nonlinearity $g(s)=\lambda s$ corresponds to the so-called
{\it Soler model of extended fermions \cite{Soler, BCDM}}.
In the general case of $g(s)$, Eqn~(\ref{4.1})
is often called the {\it generalized Soler model} (see \cite{BD, ES}).
%% The review of models of extended particles by means of nonlinear Dirac
%% fields can be found in \cite{Ran}.
\medskip

The stationary solutions of nonlinear Dirac equation are considered
 as particle-like solutions.
%--------------------------------------------------
They are the solutions of a form
$\psi_0(t,x)=e^{-i\omega t}\varphi(x)$, where
$\varphi$ is non-zero localized solution of
the stationary nonlinear Dirac equation (\ref{4.4}), see Definition \ref{def2} below.

Denote by $\psi_{\bf v}(t,x)$ {\it the moving solitary waves}
with velocity ${\bf v}\in\R^3$, $|{\bf v}|<1$,
$$
\psi_{\bf v}(t,x)=S(\Lambda_{\bf v})
\psi_0\left(\Lambda^{-1}_{\bf v}(t,x)\right),\quad x\in\R^3,\quad t\in\R,
$$
where $\Lambda_{\bf v}$  is a Lorentz transformation  (see formula (\ref{3.4}) below),
$S(\Lambda_{\bf v})$ is a matrix defined in (\ref{S}).
Put $G(s)=\int_0^s g(p)\,dp$.
The energy  functional is  given by
\be\la{HD}
{\cal E}(\psi)=%\frac12
\int_{\R^3} \left(-i\psi^*\al\cdot\nabla\psi+m\bar\psi\psi-
G(\bar\psi\psi)\right)\,dx. %%\quad Q(\psi)=\int \psi^*\psi\,dx.
\ee
%%%%%%%%%%%%%%%%%%%%%%%%%%%%%%%%%%%%
Using equalities (\ref{Pmat}), it is easily to check that
${\cal E}(\psi(t,\cdot))=\const$.
%%------------------------------------------

Our main objective is to prove that the energy-momentum relation coincides
with one of relativistic point particle, namely,
\be\label{Ein}
{\cal E}(\psi_{\bf v})=\gamma{\cal E}(\psi_0),\quad
\gamma=(1-|{\bf v}|^2)^{-1/2}.
\ee

The paper is organized as follows. In Sections 2 and 3, we check (\ref{Ein})
for nonlinear Dirac  equations (\ref{4.1}).
Section 4 concerns the Dirac equations in $\R^1$.
For (MD) and (KGD) equations, the result is obtained in Sections 5 and 6,  respectively.
%%%%%%%%%%%%%%%%%%

%%---------------------------------------------
\setcounter{equation}{0}
\section{ Standing solitary waves for Dirac equations}\label{Sec2}
%%-------------------------------------------

\begin{definition}\label{def2}
The stationary states or localized solutions of Eqn~(\ref{4.1})
are the solutions  of the form
$\psi_0(t,x)=e^{-i\omega t}\varphi_\omega(x)$, $\omega\in\R$,
such that $\varphi_\omega\in H^1(\R^3;\CC^4)$,
 and $\varphi\equiv\varphi_\omega$ is a nonzero
localized solution of the following stationary nonlinear Dirac equation
\be\label{4.4}
i\al\cdot\nabla\varphi+\omega\varphi-m\beta\varphi
+g(\bar\varphi\varphi)\beta\varphi=0,\quad x\in\R^3.
\ee
\end{definition}
%%%%%%%%%%%%%%%%%%%%%%%%%%%%%%%%%%%%%%%%%%%

The existence of solutions of Eqn~(\ref{4.4})
has been proved in \cite{BCDM, BCV, CV,ES, Merle}
under some restrictions on $G$ for $\omega\in(0,m)$.
In \cite{ES}, the following conditions  were imposed.\\
{\bf G1}.  $G\in C^2(\R;\R)$\\
 {\bf G2}. For any $s\in\R$, $g(s) s\ge \theta G(s)$
 with some $\theta>1$, ($g(s)=G'(s)$)\\
{\bf G3}. $G(0)=G'(0)=0$\\
{\bf G4}. $G(s)\ge 0$ for any $s\in\R$, and $G(A_0)>0$
 for some $A_0>0$.
%%%%%%%%%%%%%%%%%%%%%%%%%%%%%%%%%%%%%%%%%%%%%%%%%%%%%%%%
\begin{theorem} (see \cite[Theorem 1]{ES})
Let conditions {\bf G1}--{\bf G4} hold and $\omega\in(0,m)$.
Then there is an infinity of solutions of Eqn (\ref{4.4}) in
$\bigcap\limits_{2\le q<\infty}W^{1,q}(\R^3;\CC^4)$.
Each of them are critical points of the functional $I^\omega_D$,
$
I^\omega_D(\varphi)=-\ds\frac12\int\limits_{\R^3}
\left(i\varphi^*\al\cdot\nabla\varphi
-m\bar\varphi\varphi+\omega|\varphi|^2+G(\bar\varphi\varphi)\right)\,dx.
$
These solutions $\varphi\equiv\varphi_\omega$ are  of the form
(in the spherical coordinates $(r,\phi,\theta)$ of $x\in\R^3$)
\be\label{ans}
\varphi_\omega(x)=\left(\ba{c}v(r)\left(\ba{c}1\\0\ea\right)\\
iu(r)\left(\ba{c}\cos\theta\\\sin\theta e^{i\phi}\ea\right)
\ea\right),\quad
\ba {rcl}
x_1&=&r\cos\phi\sin\theta,\\
x_2&=&r\sin\phi\sin\theta,\\
x_3&=&r\cos\theta,\,\,r=|x|.
\ea\ee
Thus they correspond to classical solutions of the O.D.E. system
$$
\left\{\ba{l}
u'+\frac{2u}{r}=v[g(v^2-u^2)-(m-\omega)],\\
v'=u[g(v^2-u^2)-(m+\omega)].\ea
\right.
$$
Finally, the solutions decrease exponentially at infinity, together
with their first derivatives.
\end{theorem}
%%%%%%%%%%%%%%%%%%%%%%%%%%%%%%%%%%%%%%%5
\begin{remarks}
{\rm
(i)
Denote by ${\cal L}_D$ the Lagrangian density for considered Dirac fields,
\be\label{LD}
{\cal L}_D(\psi)=
\bar\psi(i\gamma^\mu\pa_\mu-m)\psi+G(\bar\psi\psi),
\ee
where $\gamma^\mu\pa_\mu=\gamma^0\pa_t+\gamma\cdot\nabla$
with Dirac matrices $\gamma^\mu$ ($\gamma^0=\beta$,
$\gamma^k=\beta\al_k$, $k=1,2,3$). It is easy to check that
the Euler--Lagrange equations applied to (\ref{LD})
give Eqn (\ref{4.1}). In particular, for stationary solutions $\psi_0(t,x)$,
$
{\cal L}_D(\psi_0)=
\varphi^*(\omega+i\alpha\cdot\nabla-m\beta)\varphi+G(\bar\varphi\varphi).
$
Note that $I^\omega_D(\varphi)=-(1/2)\int{\cal L}_D(\psi_0)\,dx$.
Here and below,  for simplicity,
we omit the symbol $\R^3$ in the notation of the integral
$\int_{\R^3}\dots dx$.
\medskip

(ii)
In \cite{BCV}, the existence of solutions of the form (\ref{ans})
have been proved for singular self-interactions $g(s)\sim s^{-\al}$
with some $\al\in (0,1)$.

(iii)
The stationary nonlinear Dirac equations of the form
\be\la{4.4'}
i\al\cdot\nabla\varphi+\omega\varphi-m\beta\varphi
+ \nabla F(\varphi)=0,\quad x\in\R^3,
\ee
has been studied  by Esteban and S\'er\'e in \cite{ES}.
%%If $F(\varphi)=G(\bar\varphi\varphi)$,
%%Eqn (\ref{4.4'}) coincides with (\ref{4.4}).
For a more general class of nonlinearities $F$, which do not satisfy condition
$F(\varphi)=G(\bar\varphi\varphi)$, the ansatz (\ref{ans})
is no more valid. In this case, the existence of solutions of (\ref{4.4'})
has been proved  in \cite[Theorems 2,3]{ES} with nonlinearities as
(1) $F(\varphi)=\lambda(|\bar\varphi\varphi|^{\kappa_1}+
b|\bar\varphi\gamma^5\varphi|^{\kappa_2})$ with $1<\kappa_1,\kappa_2<3/2$,
$\gamma^5=-i\al_1\al_2\al_3$, $\lambda,b>0$;
and\\ (2) $F'(0)=F''(0)=0$,
$0\le F(\varphi)\le a(|\bar\varphi|^{\kappa_3}+|\varphi|^{\kappa_4})$
with $a>0$, $2<\kappa_3\le\kappa_4<3$.
}\end{remarks}
%%%%%%%%%%%%%%%%%%%%%%%%

The following virial identity (or so-called Pokhozhaev identity \cite{Po})
was proved in \cite[Proposition 3.1]{ES}.
%%%%%%%%%%%%%%%%%%%%%%%%%%%%%%%%%%%%%%%%%%%%%%%%%%
\begin{lemma}\label{l2.8}
Let $\varphi\in H^1(\R^3;\CC^4)$ be a solution to Eqn (\ref{4.4}).
Then $\varphi(x)$ satisfies
\be\label{1.7}
i\int\varphi^*\al\cdot\nabla\varphi\,dx
=\frac32\int\left(m\bar\varphi\varphi-
\omega\varphi^*\varphi-G(\bar\varphi\varphi)\right)\,dx.
\ee
\end{lemma}
%%%%%%%%%%%%%%%%%%%%%%%%%%%%%%%%%%%%%%%%%%%%%%%%%%%%%%%%%%%%%%

Introduce the following notations
\beqn\ba{rcl}\label{notation}
I_k&\equiv& I_k(\varphi)=-i\ds\int\varphi^* \al_k \pa_k\varphi\,dy,\quad k=1,2,3,\\
Q&\equiv& Q(\varphi)=\ds\int\varphi ^*\varphi\,dx,\quad
V\equiv V(\varphi)=\int(m\bar\varphi\varphi-G(\bar\varphi\varphi))\,dx.
\ea
\eeqn
Then the equality (\ref{1.7}) is rewritten as
\be\label{vth}
\omega Q=V+\frac{2}{3}(I_1+I_2+I_3).
\ee
%%%%%%%%%%%%%%%%%%%%%
\begin{remark}
{\rm Formally, the identity (\ref{1.7}) can be proved used Derrick's technique
\cite[p.1253]{Der}. Indeed, introducing $\varphi_\lambda(x)=\varphi(x/\lambda)$
gives
\beqn
0&=&\frac{d}{d\lambda}\Big|_{\lambda=1}I^\omega_D(\varphi_\lambda)=
\frac12\frac{d}{d\lambda}\Big|_{\lambda=1}
\left[I_1(\varphi_\lambda)+I_2(\varphi_\lambda)+I_3(\varphi_\lambda)+
V(\varphi_\lambda)-\omega Q(\varphi_\lambda)\right]\nonumber\\
&=&\frac12\frac{d}{d\lambda}\Big|_{\lambda=1}
\left[\lambda^2I_1(\varphi)+\lambda^2I_2(\varphi)+\lambda^2I_3(\varphi)+
\lambda^3 V(\varphi)-\lambda^3\omega Q(\varphi)
\right]\nonumber\\
&=&I_1(\varphi)+I_2(\varphi)+I_3(\varphi)+\frac32(V-\omega Q).\nonumber
\eeqn
This gives the identity (\ref{1.7}).
The similar Derrick's technique has been used in \cite{EDKS}
for relativistic -invariant nonlinear wave equations.

Using the similar reasonings with
$\varphi_\lambda(x)=\varphi(x_1/\lambda,x_2,x_3)$,
$\varphi_\lambda(x)=\varphi(x_1,x_2/\lambda,x_3)$,
$\varphi_\lambda(x)=\varphi(x_1,x_2,x_3/\lambda)$,
it is easy to check that
\be\label{vth'}
I_1=I_2=I_3=\frac13(I_1+I_3+I_3)=\frac12(\omega Q-V).
\ee
}\end{remark}
%%%%%%%%%%%%%%%%%%%%%%%%%%%%%%%%%%%%%%
\begin{cor}\label{corDir}
Let $\varphi$  be a solution of (\ref{4.4}). Then
 the following relations hold.
\be\label{2.4}
I_1+I_2+I_3=\omega Q+\int(g(\bar\varphi\varphi)-m)\bar\varphi\varphi\,dx.
\ee
\be\label{2.5}
I_1+I_2+I_3=3\int(g(s)s-G(s))|_{s=\bar\varphi\varphi}\,dx>0.
\ee
%(ii) Let
%$$
%G'(s)s>\frac43 G(s)-\frac13ms \quad \mbox{for all }\,s>0,
%$$
\be\label{2.6}
{\cal E}_0\equiv I_1+I_2+I_3+V>0.
\ee
\end{cor}
%%------------------------
{\bf Proof}\,
 By (\ref{4.4}), we have
$$
\int\varphi^*i\al\cdot\nabla\varphi\,dx=
\int\varphi^*(-\omega\varphi+m\beta\varphi-g(\bar\varphi\varphi)\beta\varphi)\,dx.
$$
This implies the identity (\ref{2.4}). Then,
 by (\ref{vth}) and (\ref{2.4}), we obtain
$$
\omega Q= V+\frac23(I_1+I_2+I_3)=\int(m-g(\bar\varphi\varphi))\bar\varphi\varphi\,dx
+I_1+I_2+I_3.
$$
Hence,
\be\label{2.9}
\frac13(I_1+I_2+I_3)=\int(g(s)s-G(s))|_{s=\bar\varphi \varphi}\,dx.
\ee
Therefore, (\ref{2.5}) follows from (\ref{2.9}) and condition {\bf G2}, since
$$
g(s)s-G(s)>g(s)s-\theta G(s)\ge 0\quad\mbox{for all }\,s\in\R.
$$
By (\ref{HD}) and (\ref{2.4}), the energy
${\cal E}_0:={\cal E}(\psi_0(t,\cdot))$ associated with particle-like
solutions $\psi_0$ is expressed by
$$
{\cal E}_0\equiv{\cal E}_0(\varphi)=I_1+I_2+I_3+V=
\omega\int|\varphi(x)|^2\,dx+
\int(g(s)s-G(s))|_{s=\bar\varphi\varphi}\,dx>0,
%=\int(3G'(s)s-4G(s)+ms)|_{s=\bar\varphi\varphi}\,dx,
$$
by condition {\bf G2}.\bo
\medskip

Denote by $(\cdot,\cdot)$ the inner scalar product in $L^2$.
%%%%%%%%%%%%%%%%%%%%%%%%%%%%%%%%%%%%%%%%%%%%%%%%%%%%%%%%%%%%%%
\begin{lemma}\label{l2.7}
Let $\varphi$ be a solution of Eqn (\ref{4.4}),
$\varphi\in H^1(\R^3;\CC^4)$.  Then
 \be\label{4.8}
\omega(\varphi^*,\al_k\varphi)=-i(\varphi^*,\pa_k\varphi),\quad k=1,2,3.
 \ee
 \end{lemma}
 {\bf Proof}\,
  Multiply (\ref{4.4}) on the left by $\al_1$ and obtain
$$
i \pa_1\varphi+i\al_1\al_2\pa_2\varphi +i\al_1\al_3\pa_3\varphi+\omega\al_1\varphi-
m\al_1\beta\varphi+g(\bar\varphi\varphi)\al_1\beta\varphi=0.
$$
Hence
\beqn\label{4.5}
&&i(\varphi^*, \pa_1\varphi)+i(\varphi^*,\al_1\al_2\pa_2\varphi)
+i(\varphi^*,\al_1\al_3\pa_3\varphi)
+\omega(\varphi^*,\al_1\varphi)-
m(\varphi^*,\al_1\beta\varphi)\nonumber\\
&&+(\varphi^*,g(\bar\varphi\varphi)\al_1\beta\varphi)=0.
\eeqn
On the other hand, taking the adjoint of Eqn (\ref{4.4})
and multiplying on the right by $\al_1$
one obtains
$$
-i \pa_1\varphi^*-i\pa_2\varphi^* \al_2\al_1
-i\pa_3\varphi^*\al_3\al_1+\omega\varphi^*\al_1-
m\varphi^*\beta\al_1+
g(\bar\varphi\varphi)\varphi^*\beta\al_1=0.
$$
Hence,
\beqn\label{4.6}
&&-i(\pa_1\varphi^*, \varphi)-i(\pa_2\varphi^*\al_2\al_1,\varphi)
-i(\pa_3\varphi^*\al_3\al_1,\varphi)
+\omega(\varphi^*\al_1,\varphi)-
m(\varphi^*\beta\al_1,\varphi)\nonumber\\
&&+(\varphi^*\beta\al_1,g(\bar\varphi\varphi)\varphi)=0.
\eeqn
By (\ref{Pmat}), summing Eqns (\ref{4.5}) and (\ref{4.6}) gives (\ref{4.8})
for $k=1$. For $k\not=1$ the proof is similar. \bo

%%%%%%%%%%%%%%%%%%%%%%%%%%%%%%%%%%%%%%%%%%%%%%%%%%%%%%%%
\subsection{A particular ansatz  for the solutions of Eqn~(\ref{4.4})}
\label{rem2.8}
%%%%%%%%%%%%%%%%%%%%%%%%%%%%%%%%%%%%%%%

As in \cite{Lisi}, we choose to orient the angular momentum along the $x_3$-axis
and consider four families of solutions of (\ref{4.4})
which in spherical coordinates $(r,\phi,\theta)$
(i.e. $x_1=r\cos\phi\sin\theta$, $x_2=r\sin\phi\sin\theta$,
$x_3=r\cos\theta$)
are of a form
$$
\varphi^{1}(x)=\left(\ba{c}v_+(r)\left(\ba{c}1\\0\ea\right)\\
iu_+(r)\left(\ba{c}\cos\theta\\\sin\theta e^{i\phi}\ea\right)
\ea\right), \,\,\,
\varphi^2(x)=\left(\ba{c}
v_-(r)\left(\ba{c}\cos\theta\\\sin\theta e^{i\phi}\ea\right)\\
iu_-(r)\left(\ba{c}1\\0\ea\right)\\\ea
\right),
$$
$$
\varphi^3(x)=\left(\ba{c}
v_+(r)\left(\ba{c}0\\1\ea\right)\\
iu_+(r)\left(\ba{c}
\sin\theta e^{-i\phi}\\
-\cos\theta\ea\right)\\\ea
\right),\quad
\varphi^4(x)=\left(\ba{c}
v_-(r)\left(\ba{c}-\sin\theta e^{-i\phi}\\\cos\theta\ea\right)\\
iu_-(r)\left(\ba{c}0\\-1\ea\right)\\\ea
\right).
$$
If $\varphi^1,...,\varphi^4$
are substituted into Eqn (\ref{4.4}), then this equation reduces to
the following O.D.E. system for radial functions $u_\pm$ and $v_\pm$:
$$
\left\{\ba{l}
u'_\pm+\ds\frac{2u_\pm}{r}=v[g(v^2_\pm-u^2_\pm)-(m\mp\omega)],\\~\\
v'_\pm=u_\pm[g(v^2_\pm-u^2_\pm)-(m\pm\omega)].\ea
\right.
$$
%Another particular family of solutions of (\ref{4.4}) is of a form (see %\cite[p.142]{WS})
The existence of the %(classical)
solutions $u_\pm$ and $v_\pm $ %and also $\varphi^j$
follows from results  \cite{CV,Merle,BCDM, ES}.
\medskip

The total angular momentum operator is ${\bf M}={\bf L}+{\bf S}$, where
${\bf L}=x\times (-i\nabla)$ is the orbital angular momentum,
${\bf S}=\Sigma/2$
is the spin angular momentum,
$\Sigma=\left(\ba{cc}\sigma&0\\0& \sigma\ea\right)$.
 %Here ${\bf M}=(M_1,M_2,M_3)$,
%${\bf L}=(L_1,L_2,L_3)$, $x=(x_1,x_2,x_3)$,
%$\sigma=(\sigma_1,\sigma_2,\sigma_3)$.
In particular, in the spherical coordinates,
the third component of ${\bf L}$ is $L_3=-i\pa_{\phi}$.
It is easy to check the following properties of
 $\varphi^a$, $a=1,2,3,4$.
 \medskip\\
 (i) $\varphi^a$ are eigenfunctions
of the third component of ${\bf M}$ with eigenvalue $m_3=\pm1/2$.
More exactly,  $M_3\varphi^a=1/2\varphi^a$ for $a=1,2$,
$M_3\varphi^a=-1/2\varphi^a$ for $a=3,4$.
Since  ${\bf M}\varphi^1=(1/2)(\varphi^3,i\varphi^3,\varphi^1)$,
${\bf M}\varphi^2=(1/2)(-\varphi^4,-i\varphi^4,\varphi^2)$,
${\bf M}\varphi^3=(1/2)(\varphi^1,-i\varphi^1,-\varphi^3)$,
${\bf M}\varphi^4=(1/2)(-\varphi^2,i\varphi^2,-\varphi^4)$,
then $M^2_k\varphi^a=1/4\varphi^a$ for all $k=1,2,3$,
and
${\bf M}^2\varphi^a=3/4\varphi^a=j(j+1/2)\varphi^a$ for all $a$.
%where ${\bf M}^2=M_1^2+M_2^2+M_3^2$.
Hence, the quantum number $j=1/2$ for all $\varphi^a$.
%%---------------------------------
%Note that the total angular moment is equal to
%$m_z=1/2$ for $\varphi^{j}$ with $j=1,2$,
% and $m_z=-1/2$ for $\varphi^j$ with $j=3,4$.
%----------------------------------------------
\medskip\\
(ii) For the "spin--orbit" operator
${\bf K}=\beta\Sigma\cdot{\bf M}-1/2\beta
=\beta(\Sigma\cdot{\bf L}+1)$ (see \cite[p.19]{Dyson}),
we have
${\bf K}^2={\bf M}^2+1/4$. Then the eigenvalues of ${\bf K}$ are $\kappa=\pm(j+1/2)$.
Hence, for all $a$, $\varphi^a$ are eigenfunctions
of ${\bf K}$ with eigenvalues $\kappa=\pm1$,
where the quantum number $\kappa=1$ for $a=1,3$ and $\kappa=-1$ for $a=2,4$.
\medskip\\
(iii) For any solution $\varphi$ from the four families
$\{\varphi^1,\dots,\varphi^4\}$, the following equalities hold.
At first,
 $\varphi^*(x)\al_3\varphi(x)\equiv0$. Secondly,
%Hence,  by Lemma \ref{l2.7},
%\be\label{2.12}
%\int (\varphi^j(x))^*\pa_3\varphi^j(x)\,dx=0,\quad \forall j.
%\ee
%Also, for the solutions $\varphi^j$, we  have
%$\ds\int(\varphi^j(x))^*\al_k\varphi^j(x)dx=0$ for all $k,j$.
%Hence,
%%--------------------------------------
\beqn\label{2.13}
\int\varphi^*(x)\nabla\varphi(x)\,dx=0,\\
\label{2.14}
\int\varphi^*(x)\al_k\,\pa_l\varphi(x)\,dx=0\quad \mbox{for any }\,\, k\not=l.
\eeqn
%%--------------------------------------
(iv) For stationary states $\psi_0(t,x)=e^{-i\omega t}\varphi(x)$
with $\varphi$ from these particular families of solutions we have
%%%----------------------------
%V(\varphi^j)&=&4\pi\int\limits_0^{+\infty}
%\left(ms-G(s)\right)|_{s=v^2_\pm-u^2_\pm}r^2\,dr,\quad
%I_k&=&\frac{4\pi}{3}\int\limits_{0}^{+\infty}
%\left(v_\pm\left(u'_\pm+\frac{2u_\pm}{r}\right)-u_\pm v'_\pm\right)r^2\,dr\nonumber\\
%&=&\frac{4\pi}{3}\left(
%\omega\int\limits_{0}^{+\infty}
%\left(v^2_\pm+u^2_\pm\right)r^2\,dr+
%\int\limits_{0}^{+\infty}(g(s)s-ms)\Big|_{s=v^2_\pm-u^2_\pm}r^2\,dr\right),\quad k=1,2,3,\nonumber\\
$Q(\psi_0)=
4\pi\int\limits_0^{+\infty}\left(v^2_\pm+u^2_\pm\right)r^2\,dr$,
$$
{\cal E}_0\equiv{\cal E}(\psi_0)=
4\pi\omega\int\limits_{0}^{+\infty}
\left(v^2_\pm+u^2_\pm\right)r^2\,dr+4\pi
\int\limits_{0}^{+\infty}(g(s)s-G(s))\Big|_{s=v^2_\pm-u^2_\pm}r^2\,dr,
$$
where $v_\pm=v_\pm(r)$, $u_\pm=u_\pm(r)$.
Moreover, the current ${\bf J}(x)=\psi_0^*(t,x)\al\psi_0(t,x)$ equals
$$
{\bf J}(x)=4\kappa_\pm m_3u_\pm v_\pm (-\sin\phi,\cos\phi,0),
$$
where the quantum numbers $m_3=\pm1/2$, $\kappa_\pm=\pm1$
are introduced above.

%%%%%%%%%%%%%%%%%%%%%%%%%%%%%%%%%%%%%%%%%%%%%%%%%%%%

%%---------------------------------------------
\setcounter{equation}{0}
\section{ Moving solitary waves for nonlinear Dirac equations}\label{Sec3}
%%-------------------------------------------
As shown, e.g., in \cite{BD,Dyson,Th}, the Dirac equation (\ref{4.1})
with $g\equiv 0$ is Lorentz invariant.
Namely, let $\Lambda=(\Lambda_{\mu\nu})_{\mu,\nu=0}^4$ be a Lorentz transformation
and $\psi(t,x)$ be a solution of (\ref{4.1}) with $g\equiv0$.
 Then there exists a matrix
$S\equiv S(\Lambda)$ such that $\psi'(t',x')=S(\Lambda)\psi(t,x)$
satisfies the same equation in the terms of the new variables $(t',x')=\Lambda(t,x)$.
It requires the following conditions on
$S\equiv S(\Lambda)$:
\be\label{cond1S}
\al_\mu =\sum\limits_{\nu=0}^4\beta S\beta
\Lambda_{\mu\nu}\al_\nu S^{-1}\quad \mbox{with }\,\, \al_0\equiv I,
\ee
 or $S^{-1}\gamma^{\nu}S=
\sum\limits_{\mu=0}^3\Lambda_{\nu\mu}\gamma^\mu$, $\nu=0,1,2,3$,
where $\gamma^0:=\beta$, $\gamma^k:=\beta\al_k$, $k=1,2,3$
(see, e.g., \cite{BD}).
Here and below by $I$ we denote the unit $4\times4$ (or $2\times 2$)
matrix.
The nonlinear equation (\ref{4.1}) is Lorentz invariant,
if condition (\ref{cond1S}) holds and
\be\label{cond2S}
S^*\beta S=\beta.
\ee
The conditions (\ref{cond1S}) and (\ref{cond2S}) can be rewritten
in the form (cf formulas (23) and (27) from \cite{Dyson})
\be\label{condS}
S^*\beta S=\beta,\quad S^*\al_\mu S=\sum\limits_{\nu=0}^4
\Lambda_{\mu\nu}\al_\nu\quad \mbox{with }\,\, \al_0\equiv I.
\ee
The existence of the matrix $S$ satisfying conditions (\ref{condS})
follows from Pauli's Fundamental Theorem.

Let $\Lambda_{\bf v}:\R^4\to\R^4$ be a Lorentz
 transformation  (boost) with  velocity ${\bf v}\in\R^3$, $|{\bf v}|<1$:
$ \Lambda_{\bf v}(t,x)=
\Big(\gamma(t+{\bf v} \cdot x),\gamma(x^\Vert+{\bf v}t)+x^\perp\Big)$,
where $x^\Vert+x^\perp=x$, $x^\Vert\Vert {\bf v}$,
$x^\perp \perp {\bf v}$,
$ \gamma=(1-{\bf v}^2)^{-1/2}$.  Hence
  (see, e.g., \cite[formula (2.14)]{Th}),% p.45
\be\label{Lambda}
\Lambda_{\bf v}=\left(\ba{cc}
\gamma&\gamma {\bf v}^T\\
\gamma {\bf v}&I+\frac{\gamma-1}{|{\bf v}|^2}{\bf v}{\bf v}^T\ea\right),
\quad \mbox{where }\,{\bf v}{\bf v}^T=(v_iv_j)_{i,j=1}^3,
\ee
i.e.,
\be\label{3.4}
\Lambda_{\bf v}(t,x)=
\Big(\gamma(t+{\bf v}\cdot x),x+(\gamma-1){\bf v}\frac{x\cdot{\bf v}}{|{\bf v}|^2}
+\gamma{\bf v}t\Big),\quad (t,x)\in\R^4.
\ee
Note that $\det \Lambda_v=1$ and
 $\Lambda_{\bf v}^{-1}=\Lambda_{-{\bf v}}$.
The matrix $S_{\bf v}\equiv S(\Lambda_{\bf v})$ can be chosen as
\be\la{S}
S_{\bf v}=\sqrt{\frac{\gamma+1}{2}}\left(I+\al\cdot {\bf v}
\frac{\gamma}{\gamma+1}\right)=
\exp\left(\frac{\xi}2 \frac{\al\cdot {\bf v}}{|{\bf v}|}\right),
\ee
where ${\rm ch}({\xi/2})=\sqrt{(\gamma+1)/2}$ or
${\rm th}(\xi)=|{\bf v}|$.
It is easy to verify that
\be\label{4.17}
\ba{cc}
S_0=I,\quad S^*_{\bf v}=S_{\bf v},\quad
S_{-{\bf v}}=S_{\bf v}^{-1},\quad
S^2_{\bf v}=\gamma(\al\cdot {\bf v}+I),\\
S^*_{\bf v}\beta S _{\bf v}=\beta,\quad
 S^*_{\bf v}\al_j S_{\bf v}
 =\al_j+\gamma v_j I+v_j\frac{\gamma-1}{|{\bf v}|^2}\,\al\cdot{\bf v},
 \quad j=1,2,3,
\ea
\ee
and conditions (\ref{condS}) hold. In particular,
\be\label{3.5}
\gamma ( I-\al\cdot {\bf v})S_{\bf v}=S^{-1}_{\bf v},\quad
\al\cdot S_{\bf v}\left(\nabla\varphi
+{\bf v}\frac{\gamma-1}{|{\bf v}|^2}\nabla\varphi\cdot{\bf v}\right)
-\gamma S_{\bf v}\nabla\varphi\cdot {\bf v}=S_{\bf v}^{-1}\al\cdot\nabla\varphi.
%S^*_{\bf v}(\al\cdot {\bf v})S_{\bf v}
% =\gamma(\al\cdot {\bf v}+{\bf v}^2I),
 \ee
%%-------------------------------------------

Let $\omega\in(0,m)$, and  $\psi_0(t,x)=e^{-i\omega t}\varphi(x)$,
be a standing solitary wave. By $\psi_{\bf v}(t,x)$ we denote
 a {\it (moving) solitary wave} with velocity ${\bf v}\in\R^3$,
$|{\bf v}|<1$:
$$
\psi_{\bf v}(t,x)=
S_{\bf v}\psi_0(\Lambda_{\bf v}^{-1} (t,x)).
$$
In other words,
\be\label{3.6}
\psi_{\bf v}(t,x)%=S_{\bf v}\psi_0(\gamma(x^\Vert-{\bf v}t)+x^\perp,\gamma(t-{\bf v}\cdot x))
=e^{-i\omega\gamma(t-{\bf v}\cdot x)}S_{\bf v}
\varphi\Big(x+(\gamma-1){\bf v}\frac{x\cdot{\bf v}}{|{\bf v}|^2}
-\gamma{\bf v}t\Big).
\ee
%%%%%%%%%%%%%%%%%%%%%%%%%%%%%%%%%5
\begin{remark}\label{rem3.1}
 (i)
Let $\varphi(x)$ be a non-zero solution of Eqn (\ref{4.4}).
Then solitary waves $\psi_{\bf v}(t,x)$
satisfies Eqn~(\ref{4.1}). This follows from (\ref{3.5}) and (\ref{3.6}).
Indeed, substituting $\psi_{\bf v}(t,x)$ in Eqn~(\ref{4.1})
and using (\ref{4.17}) and (\ref{3.5}) we obtain
\beqn
&& \Big(i\pa_t+i\al\cdot\nabla
-m \beta+ g(\bar\psi_{\bf v}\psi_{\bf v})\beta\Big) \psi_{\bf v}(t,x)
=e^{-i\omega\gamma(t-{\bf v}\cdot{\bf x})}\Big[
\omega\gamma (I-\al\cdot {\bf v})S_{\bf v}\varphi(y)+\nonumber\\
%%%%%--------------------------------------------------------
%%=e^{-i\omega\gamma(t-{\bf v}\cdot{\bf x})}\Big[
%%\omega\gamma S_{\bf v}\varphi
%%-i \gamma S_{\bf v}\nabla\varphi\cdot {\bf v}-
%%-\omega\gamma \al\cdot{\bf v}S_{\bf v}\varphi+i\al\cdot S_{\bf v}
%%\left(\nabla\varphi
%%+{\bf v}\frac{\gamma-1}{|{\bf v}|^2}\nabla\varphi\cdot{\bf v}\right)
%%-m \beta S_{\bf v}\varphi+g(\bar\varphi\varphi) \beta S_{\bf v}\varphi\Big]\nonumber\\
%%-------------------------------------------------------
&&+i
\Big(\al\cdot S_{\bf v}\Big(\nabla\varphi
+{\bf v}\frac{\gamma-1}{|{\bf v}|^2}\nabla\varphi(y)\cdot{\bf v}\Big)
-\gamma S_{\bf v}\nabla\varphi(y)\cdot {\bf v}\Big)
-m \beta S_{\bf v}\varphi(y)+g(\bar\varphi\varphi)
\beta S_{\bf v}\varphi\Big]\nonumber\\
&=&e^{-i\omega\gamma(t-{\bf v}\cdot{\bf x})}S_{\bf v}^{-1}\Big[
\omega+i\al\cdot\nabla-m\beta+g(\bar\varphi\varphi)
\beta\Big]\varphi(y)=0,\nonumber
\eeqn
with $y=x+{\bf v}(\gamma-1){x\cdot{\bf v}}/|{\bf v}|^2-\gamma{\bf v}t$.
\medskip\\
(ii) Let $\psi'(t',x')=S(\Lambda)\psi(t,x)$,
where $\Lambda$ is a Lorentz transformation.
Denote by $J(t,x)$ the 4-current, $J^\mu(t,x)=\psi^*(t,x)\al_\mu\psi(t,x)$
(with $\al_0\equiv I$) and let $J'^\mu(t,x)=\psi'(t,x)^*\al_\mu\psi'(t,x)$.
 Then
$$
J'(t',x')=\Lambda J(t,x),
\quad \mbox{where }\,(t',x')=\Lambda(t,x).
$$
In particular, if $\Lambda$ is a boost, i.e., $\Lambda=\Lambda_{\bf v}$
with ${\bf v}\in\R^3$, and $\psi=\psi_{\bf v}$ with
$\psi_{\bf v}$ from (\ref{3.6}), then
\be\label{3.9}
J_{\bf v}(t,x)=\Lambda_{\bf v} J_0(y)
=\Lambda_{\bf v} \left(\ba{c}
\varphi^*(y)\varphi(y)\\\varphi^*(y)\al\varphi(y)\ea\right),
\ee
where $J^\mu_{\bf v}=\psi_{\bf v}^*\al_\mu\psi_{\bf v}$,
$y=x+{\bf v}(\gamma-1){x\cdot{\bf v}}/|{\bf v}|^2-\gamma{\bf v}t$.
\end{remark}
%%-----------------------------------------------

For simplicity, put  ${\bf v}=(0,0,v)\in\R^3$.
In this case we denote by $\Lambda_v$ the Lorentz transformation (boost) $\Lambda_{\bf v}$:
\be\label{3.8}
\Lambda_v:(t,x)\to \left(\gamma(t+vx_3),x_1,x_2,\gamma(x_3+vt)\right),
\quad |v|<1;
\ee
the solitary waves $\psi_v(t,x):=\psi_{\bf v}(t,x)|_{{\bf v}=(0,0,v)}$ are
\be\label{soliton}
\psi_v(t,x)=S_v\psi_0(\Lambda^{-1}_v(t,x))=
e^{-i\omega(t-vx_3)\gamma}S_v\varphi(x_1,x_2,\gamma(x_3-vt));
\ee
 the matrix $S_v$ ($S_v:=S_{\bv}$ if $\bv=(0,0,v)$) is defined as
\be\label{Sv}
S_v=\sqrt{\frac{\gamma+1}{2}}\left(I+\al_3\frac{v\gamma}{\gamma+1}
\right)=
\sqrt{\frac{\gamma+1}{2}}\left(\ba{rcl}I& \frac{v\gamma}{\gamma+1}\sigma_3 \\
\frac{v\gamma}{\gamma+1}\sigma_3 &I\ea\right),\quad v\in\R^1.
\ee
Using the explicit formulas (\ref{Sv}), we obtain the following properties
of $S_v$ (cf (\ref{4.17}), (\ref{3.5})).
\beqn\label{3.7}
\ba{c}S_0=I,\quad  S_{v}^*=S_{v},\quad S_{-v}=S_{v}^{-1},
\quad S_v^*\beta S_v=\beta,\\
S_v^*\al_3S_v=\gamma(vI+\al_3),\,\,\,
S_v^*S_v=\gamma(v\al_3+I),\,\,\,
S_v^*\al_k S_v=\al_k,\,\, k=1,2.
\ea\eeqn
In particular,
$\gamma S_v^{*}(\al_3-vI)S_v=\al_3$,
$\gamma S_v^{*} (I-\al_3v)S_v=I$.
%%%%%-----------------------------------------------------
\medskip

Given ${\bf v}=(v_1,v_2,v_3)\in\R^3$,
we impose the following conditions on $\varphi(x)$.\smallskip\\
{\bf C1}\,  $\int\varphi^*\nabla\varphi\,dx\cdot{\bf v}=0$.\smallskip\\
{\bf C2}\,
 $\sum\limits_{k,j:\,k\not=j}v_kv_j\int\varphi^*\al_k\pa_j\varphi\,dx=0$.
%%---------------------------------------------
\begin{theorem}
Let ${\bf v}\in\R^1$ with $|{\bf v}|<1$,  $\psi_{\bf v}(t,x)$
be a solitary wave of the form (\ref{3.6}),
and $\varphi$ satisfy  conditions {\bf C1} and {\bf C2}.
 Then
\be\label{th}
{\cal E}_{\bf v}=\gamma {\cal E}_0,
\ee
where ${\cal E}_{\bf v}:={\cal E}(\psi_{\bf v})$.
\end{theorem}
%%--------------------------------------------
{\bf Proof}\,
We first consider the particular case ${\bf v}=(0,0,v)\in\R^3$
when $\psi_v(t,x)$ is defined in (\ref{soliton}).
Substitute the function $\psi_v$
into (\ref{HD}) and apply equalities (\ref{3.7}):
\beqn
{\cal E}_{\bf v}&:=&
%\frac12
\int\left(-i\sum\limits_{k=1}^2
\varphi^* S^*_v\al_k S_v\pa_k\varphi
-i\varphi^* S_v^*\al_3 S_v\gamma(i\omega v\varphi+\pa_3\varphi)
+m\varphi^* S^*_v\beta S_v\varphi-G(\bar\varphi\varphi)\right)\,dx\nonumber\\
&=&%\frac{1}{2}
\int\left(
\gamma^2\varphi^* (v+\al_3)(\omega v\varphi-i\pa_3\varphi)
-i\sum\limits_{k=1}^2
\varphi^* \al_k \pa_k\varphi+m\bar\varphi\varphi-G(\bar\varphi\varphi)
\right)\,dx\nonumber,
\eeqn
where $\varphi\equiv \varphi(x_1,x_2,\gamma(x_3-vt))$.
 Changing variables
$x=(x_1,x_2,x_3)\to y:=(x_1,x_2,\gamma(x_3-vt))$,
we obtain
$$
{\cal E}_{\bf v}=%\frac12\left(
\omega \gamma v(\varphi^*, (v+\al_3)\varphi)
-{i\gamma v}(\varphi^*, \pa_3\varphi)+\gamma I_3
+\frac1{\gamma}(I_1+I_2)+\frac1{\gamma}V.
$$
%where $(\cdot,\cdot)$ denotes the scalar product in $L^2(\R^3)$.
In particular,
\be\label{E0}
{\cal E}_0 \equiv{\cal E}(\psi_0)=%\frac{1}{2}\left(
I_1+I_2+I_3+V=3I_3+V,
\ee
%%-----------------------------------------------------
since $I_1=I_2=I_3$.
Applying equalities (\ref{vth}) and (\ref{4.8}), one obtains
$$
\omega(\varphi^*, (v+\al_3)\varphi)
=v\,\omega Q+\omega(\varphi^*, \al_3\varphi)=
v(V+2I_3)-i(\varphi^*,\pa_3\varphi).
$$
 Therefore,
\beqn
{\cal E}_{\bf v}&=&
%\frac12\left[
v\gamma\Big(vV+2vI_3-i(\varphi^*,\pa_3\varphi)\Big)
-i\gamma v(\varphi^*, \pa_3\varphi)
+\gamma I_3+\frac2{\gamma}I_3+\frac1{\gamma}V\nonumber\\
&=&\gamma {\cal E}_0-2\gamma vi(\varphi^*,\pa_3\varphi).\nonumber
\eeqn
Hence identity (\ref{th}) holds iff  $(\varphi^*,\pa_3\varphi)=0$ what follows from condition {\bf C1}.

In the general case of ${\bf v}=(v_1,v_2,v_3)\in\R^3$,
we substitute $\psi_{\bf v}$ from (\ref{3.6}) in (\ref{HD}),
apply equalities (\ref{4.17}), change variables
$x\to y:=x+{\bf v}(\gamma-1)x\cdot{\bf v}/|{\bf v}|^2-\gamma{\bf v} t$,
use formulas (\ref{vth}) and  (\ref{4.8}) and obtain
${\cal E}_{\bf v}=\gamma{\cal E}_0+\eta_{\bf v}$, where, by definition,
\be\label{etav}
\eta_{\bf v}:=-2i\gamma(\varphi^*, \nabla\varphi)\cdot{\bf v}-i\gamma
\sum\limits_{j,k:j\not= k}(\varphi^*, \al_k \pa_j\varphi)v_k  v_j.
\ee
Hence, by conditions {\bf C1} and {\bf C2},  $\eta_{\bf v}=0$,
then identity (\ref{th}) follows.\bo
%%%%%%%%%%%%%%%
\bigskip

Let ${\bf v}\in\R^3$, $\Lambda_{\bf v}$ be defined in (\ref{3.4}),
$\psi_{\bf v}$ be of the form (\ref{3.6}). Write
$P_{\bf v}:=P(\psi_{\bf v})$, where $P(\psi)$ stands for the momentum operator,
$$
P(\psi):=-i\int_{\R^3}\psi^*(t,x)\nabla\psi(t,x)\,dx.
$$
To prove the next result for $P_{\bf v}$ we impose
conditions {\bf C1'} and {\bf C2'} which are stronger than conditions
{\bf C1} and {\bf C2}.\\
{\bf C1'} $\int\varphi^*\nabla\varphi\,dx=0$.\\
{\bf C2'} Let ${\bf v}=(v_1,v_2,v_3)\in\R^3$. For any $j=1,2,3$,
 $\sum\limits_{k:k\not=j} v_k\ds\int\varphi^*\al_k\pa_j\varphi\,dx=0$.
 %%-------------------------------------------
\begin{lemma}
Let $\varphi$ be a solution to Eqn~(\ref{4.4}) and
conditions {\bf C1'} and {\bf C2'} hold.
Then
\be\label{Pv}
P_{\bf v}=\gamma {\bf v}{\cal E}_0,
\ee
where ${\cal E}_0$ is defined in (\ref{E0}).
\end{lemma}
%%--------------------------------------------------
{\bf Proof}\,
By (\ref{4.17}) and (\ref{3.6}), we have
$$
P_{\bf v}=
-i\int_{\R^3}\varphi^* \gamma(\al\cdot{\bf v}+I)
\Big(i\omega\gamma{\bf v}\varphi+\nabla\varphi
+{\bf v}\kappa\nabla\varphi\cdot{\bf v}\Big)\,dx,
\quad\mbox{with }\,\kappa:=\frac{\gamma-1}{|{\bf v}|^2},
$$
where $\varphi\equiv \varphi (y)$ with
$y:=x+{\bf v}\kappa (x\cdot{\bf v})-\gamma{\bf v} t$.
Since %$\left\Vert\ds\frac{\pa y_i}{\pa x_j}\right\Vert=\gamma$,
$dy=\gamma dx$, changing variables $x\to y$ gives
$$
P_{\bf v}=
-i\int_{\R^3}\varphi^* (\al\cdot{\bf v}+I)
\Big(i\omega\gamma{\bf v}\varphi+\nabla\varphi
+{\bf v}\kappa\nabla\varphi\cdot{\bf v}\Big)\,dy.
$$
Using (\ref{1.7}) and (\ref{4.8}), we obtain
$P_{\bf v}=\gamma {\bf v}{\cal E}_0-i\xi_{\bf v}$, where, by definition,
\beqn\label{xiv}
\xi_{\bf v}&=&{\bf v}\Big(\gamma+\kappa\Big)
\int\varphi^* \nabla\varphi\cdot{\bf v}\,dy
+\int\varphi^* \nabla\varphi\,dy
+{\bf v}\kappa\sum\limits_{k,j:\,k\not=j}
\int\varphi^* \al_k\pa_j\varphi\,dy\,v_k v_j\nonumber\\
&&+
\Big(\sum\limits_{k\not=1}v_k\int\varphi^* \al_k \pa_1\varphi\,dy,
\sum\limits_{k\not=2}v_k\int\varphi^* \al_k \pa_2\varphi\,dy,
\sum\limits_{k\not=3}v_k\int\varphi^* \al_k \pa_3\varphi\,dy\Big).
\eeqn
In particular, if ${\bf v}=(0,0,v)\in\R^3$,
$$
\xi_{\bf v}=\Big(\int\varphi^* (v\al_3+1)\pa_1\varphi\,dy,
\int\varphi^* (v\al_3+1)\pa_2\varphi\,dy,
\gamma (v^2+1)\int\varphi^* \pa_3\varphi\,dy\Big).
$$
By conditions {\bf C1'} and {\bf C2'},
 $\xi_{\bf v}=0$, then identity  (\ref{Pv}) holds.
\bo
%%%--------------------------------
\begin{remark}
{\rm (i) Conditions {\bf C1'} and {\bf C2'}
(and also {\bf C1} and {\bf C2})
are fulfilled with any ${\bf v}\in\R^3$
for four families of solutions considered in Section \ref{rem2.8},
see formulas (\ref{2.13}) and (\ref{2.14}).
\medskip\\
(ii)
Let $\psi_{\bf v}(t,x)$ be of the form (\ref{3.6}) and condition {\bf C1'} hold.
 Then, by (\ref{4.17}), the charge functional is
\beqn
Q(\psi_{\bf v})&=&\int\psi^*_{\bf v}(t,x)\psi_{\bf v}(t,x)\,dx=
\int\varphi^* (y)(\al\cdot {\bf v}+I)\varphi(y)\,dy=
(\varphi^*,\varphi)+(\varphi^*,\al\varphi)\cdot{\bf v}\nonumber\\
&=&\int|\varphi(y)|^2\,dy.\nonumber
\eeqn
The last equality  follows from (\ref{4.8}) and condition {\bf C1'}.
Hence, $Q(\psi_{\bf v})=Q(\psi_0)$ for any ${\bf v}\in\R^3$.
\smallskip\\
%$$
%Q(\psi_{\bf v})=
%\int\varphi^*(y)(\al\cdot{\bf v}+I)\varphi(y)\,dy=
%\int\varphi^*(y)\nabla\varphi(y)\,dy\cdot{\bf v}+
%\int|\varphi(y)|^2\,dy.
%$$
(iii)
Let $\varphi=\varphi_\omega$ be a solution of Eqn~(\ref{4.4})
from one of four families of solutions considered in Remark \ref{rem2.8}.
Then applying  the total angular momentum operator $M_3$ to $\psi_v$,
we have
$$
M_3\psi_v%=(L_3+S_3)\psi_v
=e^{-i\omega\gamma(t-vx_3)}S_vM_3
\varphi(x_1,x_2,\gamma(x_3-vt)).
$$
Hence, if $\varphi\in\{\varphi^1,\varphi^2\}$ (see Remark \ref{rem2.8}),
then $M_3\psi_v=1/2\psi_v$. For $\varphi\in\{\varphi^3,\varphi^4\}$,
 $M_3\psi_v=-1/2\psi_v$.
}\end{remark}

\newpage
%%%%%%%%%%%%%%%%%%%%%%%%%%%%%%%%%%%%%%%%%%%%%%%%%%%%%%
\setcounter{equation}{0}
\section{Solitary waves in $1+1$ dimensions }
%The massive Gross--Neveu model
%%%%%%%%%%%%%%%%%%%%%%%%%%%%%%%%%%%%%%%%%%%%%%%%%%%%%%%

We consider the nonlinear Dirac equation in $\R^1$,
\be\label{5.1}
i\dot\psi=-i\al\psi'+m\beta\psi-\beta g(\bar\psi\psi)\psi,
\quad x\in\R^1,\quad t\in\R.
\ee
Here $\psi':=\pa_x\psi$,
$\psi(t,x)\in\CC^2$, $\al=-\sigma_2$, $\beta=\sigma_3$.
%Hence,
%\beqn
%\left\{\ba{rcl}
%i\dot\psi_1&=&\pa_x\psi_2+g(|\psi_1|^2-|\psi_2|^2)\psi_1,\\
%i\dot\psi_2&=&-\pa_x\psi_1-g(|\psi_1|^2-|\psi_2|^2)\psi_2
%\ea\right.
%\eeqn
In the case when $g(s)=s$, Eqn (\ref{5.1}) is called
{\it the massive Gross--Neveu model} (or the 1D Soler model).
The stationary states or localized solutions of (\ref{5.1})
 are the solutions of the form
$\psi(t,x)=e^{-i\omega t}\varphi_\omega$, $\omega\in(0,m)$,
such that $\varphi_\omega\in H^1(\R^1)$,
 and $\varphi\equiv\varphi_\omega$ is a nonzero
localized solution of the following stationary nonlinear Dirac equation
\be\label{5.4}
i\al\varphi'+\omega\varphi-m\beta\varphi
+g(\bar\varphi\varphi)\beta\varphi=0,\quad x\in\R^1.
\ee
The solitary wave solutions have been studied, e.g., in \cite{GN, LKG}.
Write
$$
I=-i\int_{\R^1}\varphi^* \al \varphi'\,dy,\quad
Q=\int_{\R^1}\varphi ^*\varphi\,dx,\quad
V=\int_{\R^1}(m\bar\varphi\varphi-G(\bar\varphi\varphi))\,dy.\nonumber
$$
Note that
\be\label{5.5}
\omega Q=V.
\ee
This equality can be proved similarly to (\ref{vth}).

For $v\in\R^1$, $|v|<1$, introduce the ''moving solitary waves''
$$
\psi_v(t,x)=e^{-i\omega\gamma(t-vx)}S_v\varphi(\gamma(x-vt)),\quad
S_v=\sqrt{\frac{\gamma+1}2}\left(I+\al\frac{v\gamma}{\gamma+1}\right),
\quad x\in\R^1.
$$
Note that $\al^*=\al$, $\beta^*=\beta$, $\al^2=\beta^2=I$,
$\al\beta+\beta\al=0$. Hence $S_v^*\beta S_v=\beta$, $S_v^*S_v=\gamma(v\al+I)$,
$S_v^*\al S_v=\gamma(vI+\al)$.
Consider
$$
{\cal E}_v:={\cal E}(\psi_v)=
\int_{\R^1}\left(-i\psi^*_v\al\psi'_v+m\bar\psi_v\psi_v-
G(\bar\psi_v\psi_v)\right)\,dx.
$$
Using the properties $S_v$, we obtain
\beqn
{\cal E}_v&=&\int_{\R^1}\Big(-i\varphi^*S^*_v\al S_v
(i\omega\gamma v\varphi+\gamma\varphi')
+m\bar\varphi\varphi-G(\bar\varphi\varphi)\Big)
\Big|_{\varphi=\varphi(\gamma(x-vt))}\,dx\nonumber\\
&=&-i\gamma\int_{\R^1}
\varphi^*(y)(vI+\al)(i\omega v\varphi(y)+
\varphi'(y))dy+\frac1{\gamma} V.\nonumber
\eeqn
In the last integral we changed variable $x\to y=\gamma(x-vt)$.
Hence,
$$
{\cal E}_v=\gamma v^2\omega Q
+\gamma v\omega (\varphi^*,\al\varphi)-i\gamma v(\varphi^*,v\varphi')
+\gamma I+\frac1{\gamma} V.
$$
In particular,
$$
{\cal E}_0=\int_{\R^1}(-i\varphi^*\al\varphi+m\bar\varphi\varphi-
G(\bar\varphi\varphi))\,dx=I+V.
$$
We apply equalities (\ref{5.5}) and
$\omega (\varphi^*,\al\varphi)=-i(\varphi^*,\varphi')$ (cf (\ref{4.8}))
and obtain
$$
{\cal E}_v=\gamma{\cal E}_0-2i\gamma v(\varphi^*,\varphi').
$$
Assuming that $\varphi$ satisfies the property $(\varphi^*,\varphi')=0$
(cf condition {\bf C1} or {\bf C1'}),
we have \\${\cal E}_v=\gamma{\cal E}_0$.
Moreover, under the same condition on $\varphi$, one obtains
$$
P_v=-i\int_{\R^1}\psi_v^*(t,x)\psi'_v(t,x)dx=\gamma v{\cal E}_0-
 \gamma i (v^2+1)\int_{\R^1} \varphi^*(x)\varphi'(x)\,dx
=\gamma v{\cal E}_0.
$$

%%---------------------------------------------
\setcounter{equation}{0}
\section{ Maxwell-Dirac equations}
%%-------------------------------------------
%The Maxwell-Dirac system (MD) plays a central role
%in quantum electrodynamics and describes
%the interaction of an electron with its own electromagnetic field.
%The electromagnetic interaction was introduced in order to construct a
%model of extended charged fermion,

We use natural units, in which we have rescaled length and time so that
$\hbar=c=e=1$. Then, in the Lorentz gauge, the (MD) system reads
\beqn\la{1'}
&&i \dot\psi=\Phi\psi-i \al\cdot\nabla\psi-\al\cdot{\bf A}\psi+m\beta\psi,
\quad x\in\R^3,\quad t\in\R,\\
\label{2'}
&&\left.\ba{rcl}
\ddot\Phi-\Delta\Phi=4\pi\rho\\
\ddot{\bf A}-\Delta {\bf A}= 4\pi{\bf J}\ea\right|\\
\la{3'}
&&\nabla\cdot {\bf A}+\dot \Phi=0.
\eeqn
Here $\psi$ describes the charged Dirac spinor,
$\psi\equiv \psi(t,x)\in \CC^4$ for $(t,x)\in\R^3\times\R$,
${\bf A}\equiv{\bf A}(t,x)=(A^1,A^2,A^3)$ and $\Phi\equiv\Phi(t,x)$
 are the classical electromagnetic potentials,
$m>0$, $\rho\equiv\rho(t,x)$ is charge density,
% плотность электрического заряда,
${\bf J}\equiv{\bf J}(t,x)$ is electric current.  % плотность тока,
By definition,
$$
\rho=\psi^*\psi,\quad {\bf J}=\psi^*\al\psi,
$$
$\beta$, $\al=(\al_1,\al_2,\al_3)$ are Pauli--Dirac matrices.
We also introduce notation $J=(\rho,{\bf J})$
for the 4-electromagnetic current ($J^\mu=\bar\psi\gamma^\mu\psi$),
and $A=(A^\mu)=(\Phi,{\bf A})$ for the 4-potential of the electromagnetic field.
The equation (\ref{3'}) is called the Lorentz gauge condition.

%%%%-
The magnetic and electric fields %напряженность магнитного и электрического поля
${\bf H}\equiv{\bf H}(t,x)$ and ${\bf E}\equiv{\bf E}(t,x)$ are given by
\be\label{EH}
 {\bf H}=\rot {\bf A}\equiv \nabla\times {\bf A},\quad
 {\bf E}=-\dot {\bf A}-\nabla\Phi.
 \ee
 Then, by condition (\ref{3'}), equations (\ref{2'}) become classical Maxwell's equations
 of electrodynamics
$$
  \dot {\bf H}=-\rot {\bf E},\quad
  \dot {\bf E}=\rot {\bf H}- 4\pi{\bf J},\quad
\nabla\cdot {\bf E}=4\pi\rho,\quad\nabla\cdot {\bf H}=0.
 $$
  As shown, e.g., in \cite{BD,Schiff}, this model is based on the Lagrangian density
  ${\cal L}_Q ={\cal L}_D+{\cal L}_{M}+{\cal L}_I$.
 Here ${\cal L}_D$ and ${\cal L}_M$ are  Lagrangian densities
 for the free Dirac and electromagnetic fields, resp.,
 ${\cal L}_I$ is extra term describing the interaction between $\psi$
 and the electromagnetic field,
$$
 {\cal L}_D=\bar\psi(i\gamma^\mu\pa_\mu-m)\psi,\quad
{\cal L}_M=-\frac1{16\pi} F^{\mu\nu}F_{\mu\nu}=
\frac1{8\pi}({\bf E}^2-{\bf H}^2),\quad
{\cal L}_I= -J_\mu A^\mu\equiv-\rho\Phi+{\bf J}\cdot{\bf A},
$$
where $F_{\mu\nu}$ stands for the electromagnetic field tensor,
$F_{\mu\nu}:=\pa_\nu A_\mu-\pa_\mu A_\nu$,
$\pa_\mu=\pa/\pa x^\mu$, $\mu,\nu=0,1,2,3$.
Other words,
\beqn\la{Lag}
{\cal L}_Q\equiv {\cal L}_Q(\psi, A)
=\psi^*\Big(i\pa_t+i\al\cdot\nabla-\Phi+\al\cdot{\bf A}-m\beta\Big)\psi
+\frac1{8\pi}\left(|\dot {\bf A}+\nabla\Phi|^2-|\rot {\bf A}|^2\right).
\eeqn
It is easy to check that the Euler--Lagrange equations applied to (\ref{Lag})
give Eqn (\ref{1'}) and $(\pa^2_t-\Delta)A^\mu-\pa^\mu(\pa_\nu A^\nu)=J^\mu$.
Due to the Lorentz gauge (\ref{3'}), we obtain Eqn (\ref{2'}).

Since $\pa{\cal L}_Q/(\pa\dot\Phi)=0$, the Hamiltonian density  equals
 \beqn
 {\cal H}(\psi, A)&=&
 \frac{\pa{\cal L}_Q}{\pa\dot\psi}\cdot\dot\psi+
 \frac{\pa{\cal L}_Q}{\pa\dot{\bf  A}}\cdot\dot {\bf A}-{\cal L}_Q
 =
 i\psi^*\cdot\dot\psi +\frac1{4\pi}(\dot{\bf A}+\nabla\Phi)\cdot
 \dot{\bf A}-{\cal L}_Q  \nonumber\\
 &=&\psi^*\left[
\al\cdot(-i\nabla-{\bf A})+\Phi+m\beta\right]\psi+
\frac1{4\pi}{\bf E\cdot\nabla\Phi}+ \frac1{8\pi}({\bf E}^2+{\bf H}^2).\nonumber
 \eeqn
%% Due to $\frac{\pa{\cal L}}{\pa\dot\psi}=i\psi^*$,
%% $\frac{\pa{\cal L}}{\pa\dot{\bf  A}}=\dot {\bf A}+\nabla\Phi=-{\bf E}$
 Hence the energy functional of the system (\ref{1'})--(\ref{3'})
 reads (cf \cite{BD})
%%%%%%%%%%%%%%%%%%%%%%%%%%%%%%%%%%%%%%%%%%%%
%$$
%{\cal E}(t)\equiv {\cal E}(\psi(t,\cdot), A^\mu(t,\cdot))
%={\cal E}_D(\psi)+{\cal E}_{M}(A^\mu)+{\cal E}_I(\psi,A^\mu),
%$$
%\beqn
%{\cal E}_D(\psi)&\equiv&
%\int\psi^*\left[-i \al\cdot\nabla\psi+m\beta\psi\right]\,dx\\
%{\cal E}_M(A^\mu)&\equiv&
%\int[2\pi  P^2+\frac1{8\pi}(\rot {\bf A})^2- P\cdot\nabla \Phi]\,dx \nonumber\\
%&=&\frac1{8\pi}\int\left({\bf E}^2+{\bf H}^2+2{\bf E}
%\cdot\nabla\Phi\right)\,dx,\\
%{\cal E}_I(\psi,A^\mu)&\equiv&\int\left(\rho\Phi-{\bf J}\cdot{\bf A}\right)\,dx
%\eeqn
%Here $P:=\frac{1}{4\pi}\left(\dot{\bf A}+
%\nabla\Phi\right)=-\frac{1}{4\pi }{\bf E}$.
%Since $\int {\bf E}\cdot\nabla\Phi\,dx=-4\pi\int\rho\Phi\,dx$,
%%%%%%%%%%%%%%%%%%%%%%%%%%%%%%%%%%%%%%%%%%%%%%%%%%%
%
%Energy momentum tensor of the coupled (MD) system
%is given by (see \cite[p.1131]{W})
%%%%%%%%%%%%%%%%%%%%%%%%%%%%%%%%%%%%%%%%%%%%%%%%%%%%%%%%%%%
\beqn\label{Energy}
{\cal E}(t)&\equiv& {\cal E}(\psi(t,\cdot), A(t,\cdot))=
\int{\cal H}(\psi(t,\cdot), A(t,\cdot))\,dx\nonumber\\
&=&\int\psi^*\left[
\al\cdot(-i\nabla-{\bf A})\psi+m\beta\psi\right]\,dx
+\frac1{8\pi}\int({\bf E}^2+{\bf H}^2)\,dx,
\eeqn
where ${\bf E}$ and ${\bf H}$ are defined in (\ref{EH}).
Here we use the fact that
 $\int {\bf E}\cdot\nabla\Phi\,dx=-4\pi\int\rho\Phi\,dx$.
Evidently, %By direct calculation, we see that
$\dot{\cal E}(t)=0$.

%%%-------------------------------------------------
%\setcounter{equation}{0}
\subsection{Standing solitary waves}
%%%%%%%%%%%%%%%%%%%%%%%%%%%%%%%%%%%%%%%%%%%%

%The existence of soliton-like solutions of the (MD) system was proved
%by Esteban, Georgiev and S\'er\'e \cite{EGS}.
%One of the main features of these soliton solutions is that they
%ehave as relativistic particles.

Let $\omega\in(-m,m)$. Consider a {\it stationary} solution $(\psi,A)$
of system (\ref{1'})--(\ref{3'}) such that
$\psi(t,x)=e^{-i\omega t}\varphi(x)$
and  $A=(\Phi,{\bf A})$ does not depend on $t$.
Such stationary solutions we denote by $(\psi_0,A_0)$.
Substituting these solutions in system (\ref{1'})--(\ref{2'}) we obtain
\beqn\la{2.1}
&&\left(-\omega+\Phi_0-i \al\cdot\nabla-\al\cdot{\bf A}_0+m\beta\right)\varphi=0,
\quad x\in\R^3,\\
\la{2.2}
&&\left.\ba{rcl}
-\Delta \Phi_0&=&4\pi\rho_0=4\pi\varphi^*\varphi,\\
-\Delta {\bf A}_{0}^k&=&4\pi J^k_0=4\pi\varphi^*\al_k\varphi,\quad k=1,2,3\ea\right|
\eeqn
By (\ref{2.2}), $A_0^\mu=\varphi^*\al_\mu\varphi*(1/|x|)$
(with $\al_0\equiv I$), $\mu=0,1,2,3$, i.e.,
\be\label{6.4}
\Phi_0(x)=\int\frac{\rho_0(y)}{|x-y|}\,dy,\quad
{\bf A}_0(x)=\int\frac{{\bf J}_0(y)}{|x-y|}\,dy,\quad\mbox{with }\,
\rho_0=|\varphi|^2,\quad {\bf J}_0=\varphi^*\al\varphi.
\ee
Note that the Lorentz condition (\ref{3'}) becomes
\be\label{2.3}
\nabla\cdot {\bf A}_0=0,
\ee
what follows from (\ref{2.1}) and (\ref{6.4}).
Using (\ref{2.2}) and (\ref{2.3}), we rewrite the energy associated
with stationary  states $(\psi_0, A_0)$  as
\beqn\label{6.5}
{\cal E}_0&:=&{\cal E}(\psi_0, A_0)%{\cal E}_0(\varphi)
= \int\varphi^*\left[
\al\cdot(-i\nabla-{\bf A}_0)+m\beta\right]\varphi\,dx+
\frac1{8\pi}\int\left(|\nabla\Phi_0|^2+|\rot{\bf  A}_0|^2\right)\,dx\nonumber\\
&=&\int\varphi^*\left[
-i\al\cdot\nabla+m\beta\right]\varphi\,dx+
\frac1{2}\int\left(\rho_0\Phi_0-{\bf J}_0\cdot{\bf A}_0\right)\,dx.
\eeqn
The last integral in (\ref{6.5}) is
$$
\frac{1}{2}\int_{\R^3} J_\mu(x) A_0^\mu(x)\,dx
=\frac12\int_{\R^6}\frac{J_\mu(x)J^\mu(y)}{|x-y|}\,dxdy.
$$
%%-----------------------------------------
\begin{definition}\label{defst}
The stationary states or standing waves
$(\psi_0(t,x),A_0(x)):\R\times\R^3\to\CC^4\times\R^4$
are the solutions of the (MD) system of a form
\beqn\label{statMD}
\ba{rcl}\psi_0(x,t)&=&e^{-i\omega t}\varphi_\omega(x),\\
A_0^\mu(x)&=&\ds  J^\mu*\frac1{|x|}=\int_{\R^3}\frac{J^\mu(y)}{|x-y|}\,dy,
\quad \mu=0,1,2,3.
\ea\eeqn
Here $\omega\in(-m,m)$,
$(J^\mu)=(\varphi^*_\omega\al_\mu\varphi_\omega)
=(\rho_0,{\bf J}_0)$,
and $\varphi\equiv\varphi_\omega$ is a solution of (\ref{2.1}).
\end{definition}
%%---------------------------------------------------------------

The stationary solutions of the
(MD) system were studied numerically by Lisi \cite{Lisi}.
Using variational methods, Esteban Georgiev and S\'er\'e  \cite{EGS}
have proved the existence of stationary solutions
%of the form $\psi(t,x)=e^{-i\omega t}\varphi(x)$,
with $\omega\in(-m,0)$.
%%The existence of stationary states was proved by
%%Esteban,  \cite{EGS} for $\omega\in(-m,0)$.
To state this result we introduce a functional
$$
I_{Q}^\omega(\varphi):=\frac12\int{\cal L}_Q(\psi_0,A_0)\,dx=
\frac{1}{2}\int\varphi^*
\left(i\al\cdot\nabla-m\beta+\omega\right)\varphi\,dx
-\frac14\int\int\frac{ J_\mu (x)J^\mu (y)}{|x-y|}\,dxdy.
$$
Note that if $(\psi_0,A_0)$ is a solution of the (MD) system
of the form (\ref{statMD}), then (formally) $\varphi_\omega$ is
a critical point of $I_{Q}^\omega(\varphi)$.
%------------------------------------------------
\begin{theorem}\label{t6.1}
(see \cite[Theorem 1]{EGS})
For any $\omega\in(-m,0)$, there exists a non-zero critical point
$\varphi\equiv\varphi_\omega\in H^{1/2}(\R^3;\CC^4)$ of the functional
$I_{Q}^\omega(\varphi)$.
Moreover,
$\varphi_\omega$ is a smooth function of $x$ exponentially decreasing
at infinity with all its derivatives,
and
 $\psi(x,t)=e^{-i\omega t}\varphi_\omega(x)$,
$A^\mu(x,t)=J^\mu*(1/|x|)$ are the solutions of the (MD) system.
\end{theorem}
%%-------------------------------------------------

%%%----------------------------------------------------
\subsection{Virial identities}
%%%%%%%%%%%%%%%%%%%%%%%%%%%%%%%%%%%%%%%%%%%%

The following virial identity was proved in \cite[Proposition 3.1]{ES}.
%%%%%%%-----------------------------------------------------
\begin{lemma}\label{l6.2}
Let $\varphi\in H^1(\R^3;\CC^4)$ be a solution to Eqn (\ref{2.1}).
Then $\varphi(x)$ satisfies
\be\label{4.7}
i\int\varphi^*\al\cdot\nabla\varphi\,dx=\frac32\int\left(
m\bar\varphi\varphi-
\omega\varphi^*\varphi+\frac56 J_\mu(x) A^\mu(x)\right)\,dx,
\ee
where  $J_\mu A^\mu=\rho_0\Phi_0-{\bf J}_0\cdot {\bf A}_0$,
$\rho_0=|\varphi|^2$, ${\bf J}_0=\varphi^*\al\varphi$.
\end{lemma}
%%%%%%%%%%%%%%%%%%%%%%%%%%%%%%%%%%%%%%%%%%%%%%%%%%%%%%%%%%%%%%

Let functionals  $I_k(\varphi)$, $k=1,2,3$, and $Q(\varphi)$ be as in (\ref{notation}).
 Also, we put
\beqn\label{6.11}
\ba{l}
T\equiv T(\varphi)=\ds\int\Big(\rho_0(x)\Phi_0(x)-{\bf J}_0(x)\cdot{\bf A}_0(x)\Big)\,dx
=(2\pi)^{-3}4\pi\int\frac{|\hat\rho_0(k)|^2-|{\bf J}_0(k)|^2}{k^2}\,dk,\\
 m_0\equiv m_0(\varphi)=\ds m\int\bar\varphi\varphi\,dx.
 \ea
\eeqn
%%-------------------------------------------------------
\begin{remark}
{\rm
Formally, the identity (\ref{4.7}) can be proved used Derrick's technique \cite[p.1253]{Der}.
Indeed, using notations (\ref{6.11}), we rewrite $I_{Q}^\omega(\varphi)$ as
\be\label{6.8}
I_{Q}^\omega(\varphi)=\frac12\Big(\omega Q(\varphi) -m_0(\varphi)-
I_1(\varphi)-I_2(\varphi)-I_3(\varphi)\Big)-\frac14 T(\varphi),
\ee
and introduce $\varphi_\lambda(x)=\varphi(x/\lambda)$.
Then $T(\varphi_\lambda)=\lambda^5 T(\varphi)$,
$I_k(\varphi_\lambda)=\lambda^2 I_k(\varphi)$, $Q(\varphi_\lambda)=\lambda^3Q(\varphi)$,
$m_0(\varphi_\lambda)=\lambda^3m_0(\varphi)$.
Hence,
\beqn
0&=&\frac{d}{d\lambda}\Big|_{\lambda=1}I_{Q}^\omega(\varphi_\lambda)=
\frac12\frac{d}{d\lambda}\Big|_{\lambda=1}
\left[\omega Q(\varphi_\lambda)-m_0(\varphi_\lambda)-
I_1(\varphi_\lambda)-I_2(\varphi_\lambda)-I_3(\varphi_\lambda)-
\frac12T(\varphi_\lambda)\right]\nonumber\\
&=&\frac32\Big(\omega Q (\varphi)- m_0(\varphi)\Big)-
I_1(\varphi)-I_2(\varphi)-I_3(\varphi)-\frac54 T(\varphi).\nonumber
\eeqn
Hence,
\be\label{6.7}
I_1+I_2+I_3=\frac32(\omega Q-m_0)-\frac54 T,
\ee
and the identity (\ref{4.7}) holds.}
\end{remark}
%%----------------------------------------------------------
\begin{cor}\label{cor5.5}
(i)
 Eqn (\ref{2.1}) implies the following equality,
$$\omega Q-m_0-(I_1+I_2+I_3)=T$$
(cf \cite[p.238]{Ab} or formula (3.10) in \cite{EGS}).
%This can be proved by a similar way as (\ref{6.7}) in item (i) substituting
%$\varphi_\lambda(x)=\lambda\varphi(x)$ in $I_{Q}^\omega$.
Hence, by (\ref{6.7}),
\be\label{6.14}
\omega Q-m_0=\frac12 T.
\ee
%%-----------------------------------------
Moreover,
 \be\label{6.10}
I_1+I_2+I_3=-\frac12 T.
\ee
In particular,
$I_{Q}^\omega(\varphi)=-\frac12(I_1+I_2+I_3)=\frac14 T$,
by (\ref{6.8}), (\ref{6.14}) and (\ref{6.10}).\medskip\\
%%--------------------------------
(ii) Using  (\ref{6.5}) and (\ref{6.11}), we rewrite  ${\cal E}_0$ as
\be\label{6.13}
{\cal E}_0=I_1+I_2+I_3+m_0+\frac12T.
\ee
%%-------------------------------------
By (\ref{6.10}) and (\ref{6.13}), we obtain
${\cal E}_0=m_0.
$\end{cor}

%%-----------------------------------
\begin{lemma}
The following identity holds,
\be\label{Ij}
I_j=\frac12(\omega Q-m_0)-\frac34 T+T_j,\quad j=1,2,3,
\ee
where
\be\label{Tj}
T_j:=4\pi(2\pi)^{-3}\int\frac{k_j^2(|\hat\rho_0(k)|^2-|\hat{\bf J}_0(k)|^2)}{k^4}\,dk
=\frac1{4\pi}\int\Big(|\pa_j\Phi_0(x)|^2-|\pa_j {\bf A}_0(x)|^2\Big)\,dx.
\ee
\end{lemma}
{\bf Proof}\,
Introduce $\varphi_\lambda(x)=\varphi(x_1/\lambda,x_2,x_3)$.
Then $I_1(\varphi_\lambda)=I_1(\varphi)$,
$I_k(\varphi_\lambda)=\lambda I_k(\varphi)$, $k=2,3$,
 $Q(\varphi_\lambda)=\lambda Q(\varphi)$,
$m_0(\varphi_\lambda)=\lambda m_0(\varphi)$, and
$T(\varphi_\lambda)=\ds(2\pi)^{-3}4\pi\lambda\int
\frac{|\hat\rho_0(k)|^2-|\hat{\bf J}_0(k)|^2}{k_1^2\lambda^{-2}+k_2^2+k_3^2}\,dk$.
Hence,
$$
0=\frac{d}{d\lambda}\Big|_{\lambda=1}I_{Q}^\omega(\varphi_\lambda)
=\frac12\Big[\omega Q-m_0-(I_2+I_3)-\frac12(T+2T_1)\Big].
$$
Therefore,
$I_2+I_3=\omega Q-m_0-(T+2T_1)/2$.
Together with (\ref{6.7}), the last identity  implies (\ref{Ij}) with $j=1$.
Similarly, introducing $\varphi_\lambda(x)=\varphi(x_1,x_2/\lambda,x_3)$
or $\varphi_\lambda(x)=\varphi(x_1,x_2,x_3/\lambda)$, we can verify
(\ref{Ij}) with $j=2,3$.\bo
%%%---------------------------------------------------
\begin{cor}
Since $T_1+T_2+T_3=T$, (\ref{Ij}) implies identity (\ref{6.7}).
Moreover, by (\ref{6.14}), we have
$I_j=- T/2+T_j,\quad j=1,2,3.
$
\end{cor}
%%--------------------------------------------

%It is easy to check the following assertion.
\begin{remark} (cf Lemma \ref{l2.8})
{\rm
 Let $\varphi$ be a solution of Eqn (\ref{2.1}). Then
$$
i\int\varphi^*(x)\nabla\varphi(x)\,dx+\omega\int\varphi^*(x)\al\varphi(x)\,dx=
\int({\bf J}_0(x)\Phi_0(x)-\rho_0(x){\bf A}_0(x))\,dx.
$$
Since $(\Phi_0,{\bf A}_0)$ is of the form (\ref{6.4}), then
\be\label{5.20}
\int{\bf J}_0(x)\Phi_0(x)\,dx=\int\rho_0(x){\bf A}_0(x)\,dx.
\ee
Hence, if  $(\varphi,\Phi_0,{\bf A}_0)$ is a solution of the system (\ref{2.1})--(\ref{2.2}),
 then (cf formula (\ref{1.7}))
\be\label{5.21}
i\int\varphi^*(x)\nabla\varphi(x)\,dx=-\omega\int\varphi^*(x)\al\varphi(x)\,dx.
\ee
}\end{remark}

%%%----------------------------------------------------
\subsection{A particular ansatz of stationary solutions}\label{Sec5.3}
%%%%%%%%%%%%%%%%%%%%%%%%%%%%%%%%%%%%%%%%%%%%

Abenda \cite[Theorem~A]{Ab} extended the results of Theorem \ref{t6.1} for $\omega\in(-m,m)$
%using cylindrical coordinates $(r,z,\phi)$
and proved the existence of
the particular ansatz of solutions to Eqn (\ref{2.1})--(\ref{2.2})
in the form
\be\label{exLisi}
\varphi_\omega(x)=\left(\ba{l}u_1(r,z)e^{i(m_3-1/2)\phi}\\
u_2(r,z)e^{i(m_3+1/2)\phi}\\
-i u_3(r,z)e^{i(m_3-1/2)\phi}\\
-i u_4(r,z)e^{i(m_3+1/2)\phi}
\ea\right),\quad \mbox{with }\,m_3=\pm \frac12,
\ee
\be\label{6.7'}
\Phi_0(x)=\Phi_*(r,z),\quad {\bf A}_0(x)=A_*(r,z)(-\sin\phi,\cos\phi,0),
\ee
where $(r,z,\phi)$ are the cylindric coordinates of $x\in\R^3$.
Moreover, $u_1,u_2,u_3,u_4,\Phi_*,A_*$ are
scalar real--valued smooth functions
exponentially decreasing at infinity with all its derivatives.
The system of equations
for $u_1,u_2,u_3,u_4,\Phi_*,A_*$ was derived by Lisi \cite{Lisi}.
%%-----------------------
\begin{remark}\label{rem5.8}
The solutions $(\varphi_\omega,\Phi_0,{\bf A}_0)$ of the form (\ref{exLisi}) and (\ref{6.7'})
have the following properties.
(i) $\varphi_\omega$
are  eigenfunctions of the third component of the total angular momentum
{\bf M} (see Section~\ref{rem2.8}) with eigenvalues $m_3=\pm1/2$.\\
  (ii) $\ds\int \varphi^*_\omega(x)\nabla\varphi_\omega(x)\,dx=0$.\\
 % $\ds\int \varphi^*_\omega(x)\al\varphi_\omega(x)\,dx=0$.\\
 (iii) $\ds\int \varphi^*_\omega(x)\al_k\partial_j\varphi_\omega(x)\,dx=0$
 for $k\not=j$, $k,j=1,2,3$.\\
  (iv)
  For $i\not=j$,
 $ \ds\int \pa_i{\bf \Phi}_0(x)\pa_j{\bf \Phi}_0(x)\,dx=0$
and  $ \ds\int \pa_i{\bf A}_0(x)\cdot\pa_j{\bf A}_0(x)\,dx=0$.\medskip\\
 (v)
  %The third component of the current ${\bf J}$ is $J^3=\varphi_\omega(x)\al_3\varphi_\omega(x)\equiv0$.
 The current
 ${\bf J}_0(x) =
 \psi_0^*(t,x)\al\psi_0(t,x)=\varphi_\omega^*(x)\al\varphi_\omega(x)
 =2(u_1u_4-u_2u_3)(\sin\phi,-\cos\phi,0)$,
the charge density is
 $\rho_0\equiv \rho_0(r,z)=u_1^2+u_2^2+u_3^2+u_4^2$.
 Moreover, by (\ref{6.7'}),
$$
{\bf E}_0(x)=-(\cos\phi\,\pa_r\Phi_*,\sin\phi\,\pa_r\Phi_*,\pa_z\Phi_*),\,\,
{\bf H}_0(x)=(-\cos\phi\,\pa_z A_*,-\sin\phi\,\pa_z A_*,\pa_r A_*+A_*/r).
$$
%In particular, for $i\not=j$,
%$\int{\bf E}_0^i(x){\bf E}_0^j(x)\,dx=0$ and
%$\int{\bf H}_0^i(x){\bf H}_0^j(x)\,dx=0$.
 %(vi) $I_1=I_2$.
 \end{remark}

%%%%%%%%%%%%%%%%%%%%%%%%%%%%%%%%%%%%%%%%%%%%%%%%%%%%%%%
%\setcounter{equation}{0}
\subsection{Moving solitary waves}
%%%%%%%%%%%%%%%%%%%%%%%%%%%%%%%%%%%%%%%%%%%%%%%%%%%%%%%

Consider {\it travelling solutions}
$(\psi_{\bf v},A_{\bf v})$, where
 $A_{\bf v}=(\Phi_{\bf v},{\bf A}_{\bf v})$,
 with velocity ${\bf v}\in\R^3$, $|{\bf v}|<1$:
\beqn\label{7.1}
\psi_{\bf v}(t,x)&=&S_{\bf v}\psi_0(\Lambda^{-1}_{\bf v}(t,x)),\nonumber\\
A_{\bf v}(t,x)&=&\Lambda_{{\bf v}} A_0(y)
\quad \mbox{with }\,
y=x+{\bf v}\frac{(\gamma-1)}{|{\bf v}|^2} x\cdot{\bf v}-\gamma{\bf v}t.
\eeqn
Here the stationary solutions
$(\psi_0,A_0)$ are introduced in  Definition~\ref{defst},
 $S_{\bf v}$ is defined in (\ref{S}),
 $\Lambda_{\bf v}$ is a Lorentz transformation  %with  velocity ${\bf v}$
defined in (\ref{Lambda}).
%$\psi_{\bf v}$ is called a {\it  travelling soliton} "surrounded"
%by the electromagnetic field $A_{\bf v}$.
It is easily to check that $(\psi_{\bf v},A_{\bf v})$ is a solution of the (MD) system.
Indeed, first, similarly to Remark \ref{rem3.1} (i), we obtain
$$
i \dot\psi_{\bf v}+i \al\cdot\nabla\psi_{\bf v}-m\beta\psi_{\bf v}=
e^{-i\omega\gamma(t-{\bf v}\cdot x)}S^{-1}_{\bf v}
\left(\omega+\al\cdot\nabla-m\beta\right)\varphi(y).
$$
Here and below $y$ stands for the expression
$y=x+{\bf v}{(\gamma-1)}x\cdot{\bf v}/{|{\bf v}|^2} -\gamma{\bf v}t$
(as in (\ref{7.1})).
On the other hand,
$S_{\bf v}(\Phi_{\bf v}(t,x)-\al\cdot{\bf A}_{\bf v}(t,x))S_{\bf v}=
\Phi_0(y)-\al\cdot{\bf A}_0(y)$, hence
$$
(\Phi_{\bf v}(t,x)-\al\cdot{\bf A}_{\bf v}(t,x))\psi_{\bf v}(t,x)=
e^{-i\omega\gamma(t-{\bf v}\cdot x)}
S^{-1}_{\bf v}(\Phi_0(y)-\al\cdot{\bf A}_0(y))\varphi(y),
$$
and Eqn (\ref{1'}) follows.
To verify Eqn (\ref{2'}), we put
$J^\mu_{\bf v}=\psi_{\bf v}\al_\mu\psi_{\bf v}$. Then,
by (\ref{7.1}), (\ref{2.2}), and Remark~\ref{rem3.1}~(ii), one obtains
$$
(\pa_t^2-\Delta)A_{\bf v}(t,x)=
\Lambda_{\bf v}(\pa_t^2-\Delta_x)A_0(y)
=\Lambda_{\bf v}(-\Delta_y A_0(y)) =4\pi \Lambda_{\bf v}J_0(y)=
4\pi J_{\bf v}(t,x),
$$
and Eqn (\ref{2'}) follows.
Moreover,
$\dot \Phi_{\bf v}(t,x)+\nabla_x\cdot{\bf A}_{\bf v}(t,x)=
\nabla_y\cdot {\bf A}_0(y)=0$, i.e., the Lorentz gauge condition (\ref{3'}) is fulfilled.
%%--------------------------------------------------------
%\beqn
%\left.\ba{rcl}
%\psi_v(t,x)&=&e^{i\omega(t-vx_3)\ga}S_v\varphi(y),\\
%A_v^1(t,x)&=&A_0^1(y),\quad
%A_v^2(t,x)=A_0^2(y)\\
%A_v^3(t,x)&=&\gamma A_0^3(y)+\ga v \Phi_0(y),\\
%\Phi_v(t,x)&=&\gamma \Phi_0(y)+\gamma v A_0^3(y).
%\ea\right|\quad \mbox{where }\,\,y=(x_1,x_2,\gamma(x_3-vt))
%\eeqn
%%-------------------------------------------------------------
\begin{remark}
Denote ${\bf E}_0=-\nabla\Phi_0$, ${\bf H}_0=\nabla\times {\bf A}_0$,
and ${\bf E}_{\bf v}=-\dot{\bf A}_{\bf v}-\nabla\Phi_0$,
${\bf H}_{\bf v}=\nabla\times {\bf A}_{\bf v}$, ${\bf v}\in\R^3$.
Then
\be\label{5.24}
\left.\ba{rcl}
{\bf E}_{\bf v}(t,x)&=&\gamma{\bf E}_0(y)-{\bf v}\kappa
{\bf v}\cdot {\bf E}_0(y)-\gamma {\bf v}\times {\bf H}_0(y)\\
{\bf H}_{\bf v}(t,x)&=&\gamma{\bf H}_0(y)-{\bf v}\kappa
{\bf v}\cdot {\bf H}_0(y)+\gamma {\bf v}\times {\bf E}_0(y)
\ea\right|\quad\kappa:=\frac{\gamma-1}{|{\bf v}|^2}.
\ee
\end{remark}
%%-----------------------------------------------------

We impose conditions {\bf C1} and {\bf C2} on $\varphi_\omega$
(see Section \ref{Sec3}).
Moreover, we assume the additional condition {\bf C0} on
$(\Phi_0,{\bf A}_0)$.\\
{\bf C0}\, For ${\bf v}=(v_1,v_2,v_3)\in\R^3$,
$$
\sum\limits_{i,j:\,i\not=j}v_iv_j\int\Big(\pa_i{\bf \Phi}_0(x)\pa_j{\bf \Phi}_0(x)-
\pa_i{\bf A}_0(x)\cdot\pa_j{\bf A}_0(x)\Big)\,dx=0.
$$
%%%--------------------------------------------------
\begin{remark}
(i) By Fourier transform and formulas (\ref{2.2}), condition {\bf C0} can be rewritten in the form
$$
\sum\limits_{i,j:\,i\not=j}v_iv_j\int\frac{k_ik_j}{k^4}
\Big(|\hat\rho_0(k)|^2-
|\hat{\bf J}_0(k)|^2\Big)\,dk=0.
$$
(ii)
Conditions {\bf C0}--{\bf C2} are fulfilled, for instance,
for the particular family of solutions considered in Section \ref{Sec5.3}
(see Remark \ref{rem5.8} (ii)--(iv)).
Obviously, conditions {\bf C0} and {\bf C2} are valid
in the particular case when ${\bf v}=(0,0,v)$.
\end{remark}
%%%-------------------------------------------------------

Put ${\cal E}_{\bf v}={\cal E}(\psi_{\bf v}, A_{\bf v})$, ${\bf v}\in\R^3$,
where ${\cal E}$ is defined in (\ref{Energy}),
$$
{\cal E}_{\bf v}=\int
\Big(\psi_{\bf v}^*(-i\al\cdot\nabla+m\beta)\psi_{\bf v}-{\bf J}_{\bf v}\cdot {\bf A}_{\bf v}\Big)\,dx
+\frac1{8\pi}\int({\bf E}_{\bf v}^2+{\bf H}_{\bf v}^2)\,dx.
$$
%%%%%---------------------------
Then the following result holds.
%%---------------------------------------------------------
\begin{theorem}
Let $(\psi_{\bf v}, A_{\bf v})$ be a  solitary wave of the (MD)
system and  conditions {\bf C0}--{\bf C2} hold.
Then the "particle-like" energy relation holds,
${\cal E}_{\bf v}=\gamma {\cal E}_0.$
\end{theorem}
%%%%%%%%%%%%%%%%%%%%%%%
{\bf Proof}\,
First we rewrite the term in ${\cal E}_{\bf v}$ corresponding to the  Dirac field,
\beqn
e_D&:=&\int\Big(\psi_{\bf v}^*(-i\al\cdot\nabla+m\beta)
\psi_{\bf v}\,dx\nonumber\\
&=&
\int\Big(-i\varphi^* S_{\bf v}\al\cdot S_{\bf v}
\Big(i\gamma\omega {\bf v}\varphi+\nabla\varphi+
{\bf v}\,\frac{\gamma-1}{|{\bf v}|^2}
 \nabla\varphi\cdot {\bf v}\Big)
 +m\varphi^* S_{\bf v}\beta S_{\bf v}\varphi\Big)\,dx
 \nonumber\\
&=&\int\Big(
 \gamma^2\omega\varphi^*[\al\cdot {\bf v}+{\bf v}^2]\varphi
-i\varphi^* \left[
\gamma^2(\al\cdot {\bf v}+1){\bf v}\cdot\nabla\varphi
+\al\cdot\nabla\varphi\right]
+m\bar\varphi\varphi\Big)\,dx, \nonumber
\eeqn
where $\varphi\equiv \varphi(y)$ with $y$ from (\ref{7.1}).
Here we apply formulas (\ref{4.17}) and (\ref{3.5}).
We change variables
$x\to y=x+{\bf v}(\gamma-1) {x\cdot{\bf v}}/{|{\bf v}|^2}-\gamma{\bf v}t$,
$dx=dy/\gamma$.  Using (\ref{5.21}), we obtain
\beqn\label{5.27}
e_D&=&
\omega\gamma(\varphi^*, [\al\cdot {\bf v}+{\bf v}^2]\varphi)
-i\gamma(\varphi^*, (\al\cdot {\bf v}+1){\bf v}\cdot\nabla\varphi)
+\frac1{\gamma}(I_1+I_2+I_3)+
\frac1{\gamma}m_0\nonumber\\
&=&
\omega Q \gamma{\bf v}^2 +\frac1{\gamma}(I_1+I_2+I_3)+\frac1{\gamma}m_0
+\gamma\sum\limits_{j=1}^3v_j^2I_j+\eta_{\bf v},
\eeqn
where $\eta_{\bf v}$ is defined in (\ref{etav}).
Applying 'virial' identities (\ref{6.7})  and (\ref{Ij}), we obtain
$$
e_D=\gamma (I_1+I_2+I_3+m_0)+\frac{1}2\gamma {\bf v}^2 T
+\gamma\sum\limits_{j=1}^3v_j^2T_j+\eta_{\bf v}.
$$
Moreover, by (\ref{6.10}),
\be\label{5.32}
e_D=
\gamma m_0-\frac1{2\gamma} T+\gamma\sum\limits_{j=1}^3v_j^2T_j+\eta_{\bf v}.
\ee
%%--------------------------------------------------
Second, we rewrite the "magnetic" term in ${\cal E}_{\bf v}$, i.e.,
the term corresponding to the interaction.
Since
$$\left.\ba{rcl}
{\bf A}_{\bf v}(t,x)&=&\gamma {\bf v} \Phi_0(y)+{\bf A}_0(y)
+{\bf v}\kappa\,{\bf A}_0(y)\cdot{\bf v} \\
{\bf J}_{\bf v}(t,x)&\equiv&\psi_{\bf v}\al\psi_{\bf v}=
\gamma {\bf v} \rho_0(y)+{\bf J}_0(y)+
{\bf v}\kappa\, {\bf J}_0(y)\cdot {\bf v}
\ea\right|\quad \kappa =\frac{\gamma-1}{|{\bf v}|^2},
$$
then, by (\ref{5.20}), we have
\beqn\label{5.33}
e_I&=&-\int {\bf A}_{\bf v}(t,x)\cdot{\bf J}_{\bf v}(t,x)\,dx\nonumber\\
&=&-\int \left(\gamma {\bf v}^2\Phi_0(y)\rho_0(y)
+\frac1\gamma{\bf A}_0(y)\cdot{\bf J}_0(y)
+\gamma({\bf v}\cdot{\bf A}_0(y))({\bf v}\cdot{\bf J}_0(y))\right)\,dy
-R_{\bf v},\,\,\,\,\,\,
\eeqn
where $R_{\bf v}$ stands for the  integral
$\ds  R_{\bf v}=2\gamma \int\rho_0(y){\bf v}\cdot{\bf A}_0(y)\,dy$.
%%%%%%%%%%%%%%%%%%%%

Further, using (\ref{5.24}),
we rewrite the energy corresponding to the electromagnetic field,
\beqn\label{em}
\ba{rcl}
e_M&:=&\ds\frac1{8\pi}\int({\bf E}_{\bf v}^2+{\bf H}_{\bf v}^2)\,dx
=\frac{4\gamma}{8\pi}\int{\bf v}\cdot({\bf H}_0\times{\bf E}_0)\,dy+\\
&&+\ds
\frac{\gamma}{8\pi}\int\Big({\bf E}_0^2+{\bf H}_0^2
+({\bf v}\times{\bf E}_0)^2+({\bf v}\times{\bf H}_0)^2
-({\bf v}\cdot{\bf E}_0)^2-({\bf v}\cdot{\bf H}_0)^2\Big)\,dy.
\ea\eeqn
Since
 $\ds\int{\bf H}_0\times{\bf E}_0\,dy=4\pi\int\rho_0{\bf A_0}\,dy$,
 the first term in the r.h.s. of (\ref{em}) equals $R_{\bf v}$.
 Using the formula $|a|^2|b|^2=(a\cdot b)^2+(a\times b)^2$ for
 $a,b\in\R^3$, the second term in $e_M$ can be rewritten as
$$
\frac{\gamma(1+{\bf v}^2)}{8\pi}\int
({\bf E}_0^2+{\bf H}_0^2)\,dy-\frac{\gamma}{4\pi}
\int\Big(({\bf v}\cdot{\bf E}_0)^2+({\bf v}\cdot{\bf H}_0)^2\Big)\,dy.
$$
Using formulas ${\bf E}_0=-\nabla\Phi_0$,
${\bf H}_0=\nabla\times {\bf A}_0$ and  (\ref{2.2}),
we obtain
\be\label{5.34}
e_M=\frac{\gamma(1+{\bf v}^2)}{2}\int
\Big(\rho_0\Phi_0+{\bf J}_0\cdot{\bf A}_0\Big)\,dy-\frac{\gamma}{4\pi}
\int\Big(({\bf v}\cdot\nabla\Phi_0)^2+({\bf v}\cdot(\nabla\times{\bf A}_0))^2\Big)\,dy +R_{\bf v}.
\ee
%$$
%\frac1{8\pi}\int({\bf E}_0^2+{\bf H}_0^2)\,dy=\frac{1}2\int(|\rho_0|^2+|{\bf %J}_0|^2)\,dx,
%$$
%%%%---------------------------------------------
Finally, substituting (\ref{5.32}), (\ref{5.33}) and (\ref{5.34}) in
${\cal E}_{\bf v}=e_D+e_I+e_M$ and using notations (\ref{6.11}) and (\ref{Tj}), we have
\beqn
{\cal E}_{\bf v}&=&\gamma m_0-\frac1{2\gamma}
\int\Big(\rho_0\Phi_0-{\bf J}_0\cdot{\bf A}_0\Big)\,dy
+\frac{\gamma}{4\pi}\sum\limits_{j=1}^3v_j^2\int(|\pa_j\Phi_0|^2
-|\pa_j{\bf A}_0|^2)\,dy+\eta_{\bf v}\nonumber\\
&&-\int \left(\gamma {\bf v}^2\Phi_0\rho_0
+\frac1\gamma{\bf A}_0\cdot{\bf J}_0
+\gamma({\bf v}\cdot{\bf A}_0)({\bf v}\cdot{\bf J}_0)\right)\,dy
-R_{\bf v}\nonumber\\
&&+\frac{\gamma(1+{\bf v}^2)}{2}\int
\Big(\rho_0\Phi_0+{\bf J}_0\cdot{\bf A}_0\Big)\,dy-\frac{\gamma}{4\pi}
\int\Big(({\bf v}\cdot\nabla\Phi_0)^2+({\bf v}\cdot(\nabla\times{\bf A}_0))^2\Big)\,dy
+R_{\bf v}\nonumber\\
&=&\gamma m_0+\eta_{\bf v}+
\frac{\gamma}{4\pi}\Big(\sum\limits_{j=1}^3v_j^2\int|\pa_j\Phi_0|^2\,dy-
\int({\bf v}\cdot\nabla\Phi_0)^2\,dy\Big)\nonumber\\
&&\!\!\!+\frac{\gamma}{4\pi}\int\!\!\Big(
4\pi{\bf v}^2{\bf A}_0\cdot{\bf J}_0-
4\pi({\bf v}\cdot{\bf A}_0)({\bf v}\cdot{\bf J}_0)-
({\bf v}\cdot(\nabla\times{\bf A}_0))^2-
\sum\limits_{j=1}^3v_j^2|\pa_j{\bf A}_0|^2\Big)dy.
 \label{5.35}
\eeqn
Using Fourier transform, relation $\hat{\bf J}_0(k)=k^2\hat{\bf A}_0(k)/(4\pi)$
and formula $|a|^2|b|^2=|a\cdot b|^2+|a\times b|^2$,
 we rewrite  the last integral in (\ref{5.35}) in the form
$$
\frac{\gamma}{4\pi}(2\pi)^{-3}\int\Big(|k\times({\bf v}\times
\hat{\bf A}_0)|^2
-\sum_{j=1}^3v_j^2k_j^2|\hat{\bf A}_0|^2\Big)\,dk.
$$
 By condition (\ref{2.3}), the last expression is
$$
 \frac{\gamma}{4\pi}(2\pi)^{-3}\int\Big((k\cdot{\bf v})^2-
\sum_{j=1}^3v_j^2k_j^2\Big)|\hat{\bf A}_0|^2\,dk=
\frac{\gamma}{4\pi}\sum\limits_{i,j:\,\,i\not=j}v_iv_j\int\pa_i{\bf A}_0(x)\cdot\pa_j{\bf A}_0(x)\,dx.
$$
Hence, by (\ref{5.35}),
${\cal E}_{\bf v}=\gamma m_0+\eta_{\bf v}+\tilde\eta_{\bf v}$,
where, by definition,
$$
\tilde\eta_{\bf v}:=-\frac{\gamma}{4\pi}\sum\limits_{i,j;\,\,i\not=j}v_iv_j
\int\Big(\pa_i{\bf \Phi}_0(x)\pa_j{\bf \Phi}_0(x)-
\pa_i{\bf A}_0(x)\cdot\pa_j{\bf A}_0(x)\Big)\,dx.
$$
%Therefore, ${\cal E}_{\bf v}=\gamma m_0=\gamma {\cal E}_0$ iff
%$\eta_{\bf v}+\tilde\eta_{\bf v}=0$.
Finally, conditions {\bf C0}--{\bf C2} yield
$\eta_{\bf v}=\tilde\eta_{\bf v}=0$.
Therefore,
${\cal E}_{\bf v}=\gamma m_0=\gamma {\cal E}_0$,
by Corollary \ref{cor5.5}~(ii).\bo
\medskip
%%%------------------------------

Denote by $P=(P^1,P^2,P^3)$ the momentum operator for the (MD) system,
%(cf \cite[Ch.1,p.39]{BSh}),
$$
P(\psi,A)=-i\int\psi^*(t,x)\nabla\psi(t,x)\,dx
+\frac1{4\pi}\int \Big(\dot\Phi(t,x)\nabla\Phi(t,x)
-\sum\limits_{k=1}^3\dot A^k(t,x)\nabla A^k(t,x)\Big)\,dx.
$$
 Put $P_{\bf v}:=P(\psi_{\bf v},A_{\bf v})$, ${\bf v}\in\R^3$.
 We impose conditions {\bf C1'} and {\bf C2}' on $\varphi_\omega$
 (see Section \ref{Sec3}).
  Moreover, we impose a stronger condition {\bf C0'} on $A_0$
  than {\bf C0}.\\
  {\bf C0'}\,
  Let ${\bf v}=(v_1,v_2,v_3)\in\R^3$. For any $k=1,2,3$,
$$
\sum\limits_{j:\,j\not=k}v_j\int\Big(\pa_j{\bf \Phi}_0(x)\pa_k{\bf \Phi}_0(x)-
\pa_j{\bf A}_0(x)\cdot\pa_k{\bf A}_0(x)\Big)\,dx=0.
$$

 Note that conditions {\bf C0'}--{\bf C2'} are fulfilled for the particular ansatz
 of solutions $\varphi\equiv\varphi_\omega$ considered in Section \ref{Sec5.3},
 see Remark \ref{rem5.8} (iv).
% since for such solutions
% we have $\hat f_\varphi(k)=\eta F_{x\to k}[v^2(|x|)-u^2(|x|)]$
% and equalities (\ref{2.13}) and (\ref{2.14})  hold.

%%----------------------------------
 \begin{lemma} Let conditions {\bf C0'}--{\bf C2'} hold. Then
 $P_{\bf v}=\gamma{\bf v}{\cal E}_{0}$.
 \end{lemma}
%%%------------------------------------------------------
{\bf Proof}\, Using (\ref{7.1}) and (\ref{4.17}), we rewrite the term
in $P_{\bf v}$ corresponding to $\psi_{\bf v}$,
\beqn
P(\psi_{\bf v})&:=&-i
\int \psi^*_{\bf v}(t,x)\nabla\psi_{\bf v}(t,x)\,dx\nonumber\\
&=&-i\int\varphi^*(y)(\al\cdot{\bf v}+I)\Big(i\omega\gamma{\bf v}\varphi(y)+
\nabla\varphi(y)+{\bf v}\kappa\nabla\varphi(y)\cdot{\bf v}\Big)\,dy,\nonumber
\eeqn
where $\kappa:=(\gamma-1)/(|{\bf v}|^2)$.
Using notations (\ref{notation}) and formula (\ref{5.21}), we have
$$
P(\psi_{\bf v})={\bf v}\gamma\omega Q+(I_1v_1,I_2v_2,I_3v_3)+
{\bf v}\kappa\sum\limits_{j=1}^3v_j^2 I_j-i\xi_{\bf v},
$$
where $\xi_{\bf v}$ is defined in (\ref{xiv}).
Applying  (\ref{Ij}) and (\ref{6.7}), we obtain
\be\label{5.36}
P(\psi_{\bf v})={\bf v}\gamma{\cal E}_0+(T_1v_1,T_2v_2,T_3v_3)+
{\bf v}\kappa\sum\limits_{j=1}^3v_j^2 T_j-i\xi_{\bf v}.
\ee
By conditions {\bf C1'} and {\bf C2}', $\xi_{\bf v}=0$.
%%-----------------------------------------------------------

Further, the second term
in $P_{\bf v}$ corresponding to $A_{\bf v}$ is
\beqn\label{5.37}
P(A_{\bf v})&:=&\frac1{4\pi}
\int \Big(\dot\Phi_{\bf v}(t,x)\nabla\Phi_{\bf v}(t,x)-
\dot{\bf A}_{\bf v}(t,x)\cdot\nabla {\bf A}_{\bf v}(t,x)\Big)\,dx\nonumber\\
&=&-\frac1{4\pi}\int\Big(({\bf v}\cdot\nabla\Phi_0(y))\nabla\Phi_0(y)+
{\bf v}\kappa({\bf v}\cdot\nabla\Phi_0(y))^2\nonumber\\
&&
-\sum\limits_{n=1}^3\Big(({\bf v}\cdot\nabla A^n_0(y))\nabla A^n_0(y)+
{\bf v}\kappa({\bf v}\cdot\nabla A^n_0(y))^2\Big)
\Big)\,dy\nonumber\\
&=&-(T_1v_1,T_2v_2,T_3v_3)-
{\bf v}\kappa\sum\limits_{j=1}^3v_j^2T_j-\tilde\xi_{\bf v},
\eeqn
where $\tilde\xi_{\bf v}$ stands for the following vector
$$
\tilde\xi_{\bf v}:=
\left(\sum\limits_{j\not=1}v_j T_{1j},
\sum\limits_{j\not=2}v_j T_{2j},
\sum\limits_{j\not=3}v_j T_{3j}\right)+
{\bf v}\kappa\sum\limits_{i,j:\,\,i\not=j}v_i v_j T_{ij}.
$$
Here by $T_{ij}$ we denote the integral
$$
T_{ij}:=\frac1{4\pi}\int\Big(\pa_i{\bf \Phi}_0(y)\pa_j{\bf \Phi}_0(y)-
\pa_i{\bf A}_0(y)\cdot\pa_j{\bf A}_0(x)\Big)\,dy.
$$
By condition {\bf C0}', $\tilde\xi_{\bf v}=0$.
Hence, formulas (\ref{5.36}) and (\ref{5.37}) yield
$$
P_{\bf v}=
P(\psi_{\bf v})+P(A_{\bf v})=\gamma{\bf v}{\cal E}_{0}.\quad\bo
$$

%%%%%%%%%%%%%%%%%%%%%%%%%%%%%%%%%%%%%%%%%%%%%%%%%%%%%%%
\setcounter{equation}{0}
\section{The  Klein--Gordon--Dirac equations }%model with Yukawa interaction
%%%%%%%%%%%%%%%%%%%%%%%%%%%%%%%%%%%%%%%%%%%%%%%%%%%%%%%
%The (KGD) system
%arises in the Yukawa model (see, i.e., \cite[v.1, Ch.10]{BD}).
We consider the Klein--Gordon--Dirac (KGD) system
arising in the Yukawa model (see, for instance, \cite[\S49]{BD})
and describes the interaction between the Dirac and scalar (or pseudoscalar) fields.
 This system is based on the Lagrangian density
\be\label{Lagr}
{\cal L}(\psi,\chi)={\cal L}_D(\psi)+{\cal L}_{KG}(\chi)+{\cal L}_I(\psi,\chi).
\ee
Here ${\cal L}_D(\psi)$ and ${\cal L}_{KG}(\chi)$ are the Lagrangian densities
for the nonlinear Dirac field $\psi$   and for the free Klein--Gordon field $\chi$,
respectively, "extra" term ${\cal L}_I$  describes the Yukawa interaction between the fields.
${\cal L}_D(\psi)$ is defined in (\ref{LD}),
$$
%\ba{rcl}
%{\cal L}_D(\psi)&=&
%i\bar\psi\gamma^\mu\pa_\mu\psi-m\bar\psi\psi+G(\bar\psi\psi),\\
{\cal L}_{KG}(\chi)=
\frac12\Big(|\dot\chi|^2-|\nabla\chi|^2-M^2\chi^2\Big),\quad
{\cal L}_I(\psi,\chi)=\eta\bar\psi\Gamma\psi\chi,
$$
where $\chi$ is a (real) scalar field, %$\psi$ is a Dirac spinor field,
 $M>0$, $\eta$ is a constant, and $\Gamma$ is some $4\times4$ matrix.
This model with $G\equiv0$ and $\Gamma=I$
has been studied by  Chadam and Glassey \cite{CG74} and Esteban {\it et al.} \cite{EGS}.
In another model presented by Ranada and Vazquez \cite{RanV}
the self--coupling is
$G(\bar\psi\psi)=\lambda(\bar\psi\psi)^2$ (as in the Soler model)
and  $\Gamma=i\gamma^5$ with
$\gamma^5=i\gamma^0\gamma^1\gamma^2\gamma^3=\left(\ba{cc}0&I\\I&0\ea\right)$.

For simplicity, we consider the case $\Gamma=I$.
Then applied the Lagrange--Euler
equations to (\ref{Lagr}) we come to the following system
\beqn
(-i\ga^\mu\pa_\mu+m-g(\bar\psi\psi))\psi&=&
\eta\chi\psi,\quad x\in\R^n,\quad t\in\R,\label{8.1}\\
(\pa_t^2-\Delta+M^2)\chi&=&\eta\bar\psi\psi,\label{8.2}
\eeqn
where $g(s)=G'(s)$, $n=1,3$ (cf \cite[p.5]{CG74}).
If $n=1$, we put $\psi=(\psi_1,\psi_2)$,
%$(\ga^0)^*=\ga^0$,
%$(\ga^1)^*=-\ga^1$,
$\beta=\sigma_3$, $\alpha=-\sigma_2$.
Below we consider the case $n=3$ only.
The case $n=1$ can be studied by a similar way.
By (\ref{Lagr}), the Hamiltonian density reads
$$
{\cal E}(\psi,\chi)=
\int\left(\psi^*(-i\al\cdot\nabla+m\beta)\psi-G(\bar\psi\psi)
+\frac1{2}
(|\dot \chi|^2+|\nabla\chi|^2+M^2|\chi|^2)-\eta\bar\psi\psi\chi\right)\,dx
$$

If $G\equiv0$,
the local existence and uniqueness of solutions to system (\ref{8.1})--(\ref{8.2})
were obtained by  Chadam and Glassey \cite{CG74}.
 The existence
for the stationary solutions was given by Esteban {\it et al.}
\cite[Th.2]{EGS} also only in the case when $G\equiv0$.
In spite of this fact
we verify below the identity (\ref{Ein}) for (KGD) system
assuming that either the self-coupling $G$ vanishes or satisfies
the conditions {\bf G1}--{\bf G4} (see Section \ref{Sec2}).

%%%------------------------
\subsection{Standing waves for the (KGD) equations}
%%--------------------------------------------------

\begin{definition}
Let $\omega\in(-m,m)$. The standing waves of the (KGD) system are
the stationary solutions $(\psi_0,\chi_0)$ of the form
\beqn\label{statKGD}
\psi_0(t,x)=e^{-i\omega t}\varphi(x),\quad
\chi_0(x)=\frac{e^{-M|x|}}{4\pi|x|}*f_\varphi,\quad
\mbox{with }\,f_\varphi:=\eta\bar\varphi\varphi,
\eeqn
where $\varphi\in H^{1}(\R^3;\CC^4)$ satisfies the following equation
\be\label{6.15}
\left(\omega+i\al\cdot\nabla-m\beta+\eta\chi_0\beta
+g(\bar\varphi\varphi)\beta\right)\varphi=0.
\ee
\end{definition}
%%------------------------------------------------

Put $I^\omega(\varphi)=\ds\frac12\int{\cal L}(\psi_0,\chi_0)\,dx$.
Then, by (\ref{Lagr}) and (\ref{statKGD}),
$$
I^\omega(\varphi)=\frac12\int\limits_{\R^3}
\Big(\varphi^*\left(i\al\cdot\nabla-m\beta+\omega\right)
\varphi+G(\bar\varphi\varphi)\Big)\,dx+\frac1{16\pi}\int\limits_{\R^6}
\frac{e^{-M|x-y|}}{|x-y|} f_\varphi(x) f_\varphi(y)\,dxdy.
$$
Note that  if $(\psi_0,\chi_0)$ is a stationary solution of the (KGD) equations,
then (formally) $\varphi\equiv\varphi_\omega$ is a critical point
of $I^\omega(\varphi)$.

%%------------------
\subsubsection{A particular family of solutions}\label{Sec6.1.1}
%%----------------------------------------------

In the spherical coordinates $(r,\phi,\theta)$ of $x\in\R^3$,
the  particular family of the stationary solutions $(\psi_0,\chi_0)$
 is given by
$$
\psi_0(t,x)=e^{-i\omega t}\varphi_\omega(x),\,\,\,
\varphi_\omega(x)=
\left(\ba{c}v\left(\ba{c}1\\0\ea\right)\\iu\left(\ba{c}\cos\theta\\
e^{i\phi}\sin\theta\ea\right)\ea\right),
\quad \ba{ll}\chi_0(x)=\left\{\ba{ll}\chi_*\cos\theta,& \mbox{if }\,\Gamma=i\gamma^5,\\
\chi_*,& \mbox{if }\,\Gamma=I,\ea\right.
\ea
$$
where $u,v,\chi_*$ being radial functions.
In the case $\Gamma=i\gamma^5$, this ansatz
has been studied numerically in \cite{RanV}.
 In the case when $\Gamma=I$,
the  functions $u,v$ are
classical solutions of the following system:
$$
\left\{\ba{l}
u'+\ds\frac{2u}{r}=v[g(v^2-u^2)-(m-\omega)+\eta\chi_*],\\
v'=u[g(v^2-u^2)-(m+\omega)+\eta\chi_*].\ea
\right.
$$
The function $\chi_*$ is a solution of the equation
$-\chi''_*-\ds\frac2{r}\chi'_*+M^2\chi_*=\eta(v^2-u^2)$ or
$\chi_*(|x|)=\ds\eta\int\frac{e^{-M|x-y|}}{4\pi|x-y|}
\Big(v^2(|y|)-u^2(|y|)\Big)\,dy$.
%%-----------------
%%The existence of stationary solutions was proved by Esteban {\it et al.}
% \cite{EGS} in the case $g\equiv0$.
%%-------------------------------------
%\begin{theorem}
%(cf \cite[Th.2]{EGS})
%Let conditions {\bf G1}--{\bf G4} hold and $\omega\in(-m,0)$. There
%is an infinity of solutions $\varphi\equiv\varphi_\omega$ of Eqn %(\ref{6.15}).
%These solutions are critical points
%of the functional $I^\omega(\varphi)$.
%They are of the form (\ref{ans}). Moreover, $\varphi_\omega$ is a smooth %function of $x$
%and it exponentially decreases at infinity with all its derivatives. Finally, $\psi=e^{-i\omega t}\varphi$,
%$\chi=\frac{e^{-M|x|}}{4\pi|x|}*f_\varphi$ are the stationary solutions of the (KGD) system.
%\end{theorem}
%%%%%%%%%%%%%%%%%%%%%%%%
%
%This theorem can be proved by a modification of the results
%of Esteban {\it et al.} \cite{EGS} for (KGE) equations with $G\equiv0$ and \cite{ES} for nonlinear Dirac equations.
%%----------------------------------------------

%------------------
\subsubsection{A virial identity}
%%----------------------------------------------
Let $I_k\equiv I_k(\varphi)$, $V \equiv V(\varphi)$,
$Q\equiv Q(\varphi)$ be as in (\ref{notation}), and
\beqn\label{9.11}
\ba{rcl}
&& R\equiv R(\varphi):=\ds\int\limits_{\R^3}
\chi_0(x)f_\varphi(x)\,dx
=\int\limits_{\R^6}\frac{e^{-M|x-y|}}{4\pi|x-y|}f_\varphi(x)f_\varphi(y) \,dxdy,\\
&& R_1\equiv R_1(\varphi):=\ds\frac1{4\pi}\int\limits_{\R^6}
e^{-M|x-y|}f_\varphi(x)f_\varphi(y) \,dxdy. \ea
\eeqn
%%--------------------------
Note that by Parseval identity and formulas (\ref{statKGD}),
\beqn\label{8.6}
\ba{rcl}
R(\varphi)&=&\ds(2\pi)^{-3}\int_{\R^3}\frac{|\hat f_{\varphi}(k)|^2}{k^2+M^2}\,dk,\\
R_1(\varphi)&=&\ds2M(2\pi)^{-3}\int_{\R^3}\frac{|\hat f_{\varphi}(k)|^2}{(k^2+M^2)^2}\,dk
=2M\int|\chi_0(x)|^2\,dx,
\ea\eeqn
where $\hat f_{\varphi}$ denotes the Fourier transform of $f_{\varphi}$.
Using (\ref{8.6}), we rewrite $I^\omega(\varphi)$   as
$$
I^\omega(\varphi)
=\frac12\left(\omega Q(\varphi)-V(\varphi)-
I_1(\varphi)-I_2(\varphi)-I_3(\varphi)+\frac12R(\varphi)\right).
$$
%%%%%%%%%%%%%%%%%%%%%%%%%%%%%%%%%%%%%%%%%%%%%%%%%%
\begin{lemma}\label{l9.1}
Let $\varphi\in H^1(\R^3;\CC^4)$ be a solution to Eqn (\ref{6.15}).
Then the following identities hold.
\be\label{8.3}
\omega Q=\frac23(I_1+I_2+I_3)+V-\frac16(5R-M\,R_1).
\ee
%\beqn\label{9.7}
%&&i\int_{\R^3}\varphi^*\al\cdot\nabla\varphi\,dx-\frac32\int_{\R^3}
%\left(m\bar\varphi\varphi-G(\bar\varphi\varphi)-
%\omega\varphi^*\varphi\right)\,dx\nonumber\\
%&&=\frac1{16\pi}\int_{\R^6}f_\varphi(x) f_\varphi(y)e^{-M|x-y|}\Big(\frac5{|x-y|}-M\Big)\,dx.
%\eeqn
Moreover,
\be\label{IjKG}
I_j(\varphi)=\frac12(\omega Q(\varphi)-V(\varphi))+\frac34R(\varphi)-
\frac{M}{4}R_1(\varphi)-P_j(\varphi),\quad j=1,2,3,
\ee
where $P_j(\varphi)$ stands for the following functional
\be\label{Pj}
P_j\equiv P_j(\varphi)=(2\pi)^{-3}\int\frac{k_j^2|\hat f_\varphi(k)|^2}{(k^2+M^2)^2}\,dk
=\int|\pa_j\chi_0(x)|^2\,dx.
\ee
\end{lemma}
%%------------------------------
\begin{remark}
{\rm
(i)
The  virial identity (\ref{8.3}) was derived in \cite{EGS} in the case when $G\equiv0$.
Formally, this identity  can be proved used Derrick's technique.
Indeed, introduce $\varphi_\lambda(x)=\varphi(x/\lambda)$.
Then
$R(\varphi_\lambda)=\lambda^5\ds\int_{\R^6}\frac{e^{-\lambda M|x-y|}}{4\pi|x-y|}
f_\varphi(x) f_\varphi(y)\,dxdy$,
$I_k(\varphi_\lambda)=\lambda^2 I_k(\varphi)$, $Q(\varphi_\lambda)=\lambda^3Q(\varphi)$,
$V(\varphi_\lambda)=\lambda^3 V(\varphi)$.
Hence,
\beqn
0&=&\frac{d}{d\lambda}\Big|_{\lambda=1}I^\omega(\varphi_\lambda)=
\frac12\frac{d}{d\lambda}\Big|_{\lambda=1}
\left[\omega Q(\varphi_\lambda)-V(\varphi_\lambda)-
I_1(\varphi_\lambda)-I_2(\varphi_\lambda)-I_3(\varphi_\lambda)+
\frac12R(\varphi_\lambda)\right]\nonumber\\
&=&\frac32\Big(\omega Q (\varphi)- V(\varphi)\Big)-
I_1(\varphi)-I_2(\varphi)-I_3(\varphi)+
\frac14( 5R(\varphi)-M\, R_1(\varphi)),\nonumber
\eeqn
and the identity (\ref{8.3}) holds.\medskip\\
%%-------------------------------------------
(ii)
Introduce
$\varphi_\lambda(x)=\varphi(x_1/\lambda, x_2,x_3)$.
Then $I_1(\varphi_\lambda)=I_1(\varphi)$,
$I_k(\varphi_\lambda)=\lambda I_k(\varphi)$, $k=2,3$,
$Q(\varphi_\lambda)=\lambda Q(\varphi)$,
$V(\varphi_\lambda)=\lambda V(\varphi)$,
%$R(\varphi_\lambda)=\lambda^2\ds\int_{\R^6}
%\frac{e^{-M|x_\lambda-y_\lambda|}}{4\pi|x_\lambda-y_\lambda|}
%f_\varphi(x) f_\varphi(y)\,dxdy$,
%where $x_\lambda=(\lambda x_1,x_2,x_3)$.
$R(\varphi_\lambda)=\ds(2\pi)^{-3}\lambda\int
\frac{|\hat f_\varphi(k)|^2\,dk}{k_1^2\lambda^{-2}+k_2^2+k^2_3+M^2}$.
By (\ref{Pj}), we have
$\ds\frac{d}{d\lambda}R(\varphi_\lambda)|_{\lambda=1}=R+2P_1$. Hence
$$
0=\frac{d}{d\lambda}\Big|_{\lambda=1}I^\omega(\varphi_\lambda)=
\frac12
\Big(\omega Q (\varphi)- V(\varphi)-I_2(\varphi)-I_3(\varphi)+
\frac12 R(\varphi)+ P_1(\varphi)\Big).
$$
Therefore,
\be\label{I2+I3}
I_2(\varphi)+I_3(\varphi)=\omega Q (\varphi)- V(\varphi)+\frac12R(\varphi)+P_1(\varphi).
\ee
Therefore,  identities (\ref{8.3}) and (\ref{I2+I3}) imply (\ref{IjKG}) with $j=1$.
Similarly, introducing
$\varphi_\lambda(x)=\varphi(x_1, x_2/\lambda,x_3)$ or
$\varphi_\lambda(x)=\varphi(x_1, x_2,x_3/\lambda)$ gives
(\ref{IjKG}) with $j=2,3$.
Note that (\ref{8.3}) follows from (\ref{IjKG}), since $P_1+P_2+P_3=R-MR_1/2$.
}\end{remark}
%%--------------------------------------------------
\begin{cor} (cf Corollary \ref{corDir}, Corollary \ref{cor5.5})
Let $\varphi$  be a solution of Eqn~(\ref{6.15}). Then
 the following relations hold.
 (i) By (\ref{6.15}),
\be\label{8.4}
I_1+I_2+I_3=\omega Q+\int(g(\bar\varphi\varphi)-m)\bar\varphi\varphi\,dx+R.
\ee
(ii) Using identities (\ref{8.3}) and (\ref{8.4}), we obtain
$$
\omega Q= \frac23(I_1+I_2+I_3)+V-\frac16(5R-MR_1)
=I_1+I_2+I_3+\int(m-g(\bar\varphi\varphi))\bar\varphi\varphi\,dx-R.
$$
Hence,
%$$
%\frac13(I_1+I_2+I_3)=\int(g(s)s-G(s))|_{s=\bar\varphi %\varphi}\,dx+\frac16(R_1+MR_1).
%$$
\be\label{8.5}
I_1+I_2+I_3=3\int(g(s)s-G(s))|_{s=\bar\varphi\varphi}\,dx+\frac12(R+MR_1)>0,
\ee
by (\ref{8.6}) and condition {\bf G2}.
\medskip\\
%%-------------------------------------------------
(iii)
The total energy associated to particle-like solutions
$(\psi_0,\chi_0)$ is
\be\label{9.14}
{\cal E}_0:={\cal E}(\psi_0,\chi_0)=I_1+I_2+I_3+V-\frac12R.
\ee
Using (\ref{8.4}), we have
$$
{\cal E}_0=
\omega\int|\varphi(x)|^2\,dx+
\int(g(s)s-G(s))|_{s=\bar\varphi\varphi}\,dx+\frac12R>0,
%=\int(3G'(s)s-4G(s)+ms)|_{s=\bar\varphi\varphi}\,dx,
$$
by condition {\bf G2}.
\medskip\\
%%------------------------
(iv)
%Denote by $(\cdot,\cdot)$ the inner scalar product in $L^2$.
 Similarly to Lemma \ref{l2.7} it can be proved
that identity (\ref{4.8}) holds.
\end{cor}
%%%-------------------------------------

%%%------------------------
\subsection{Moving waves for the (KGD) equations}
%%--------------------------------------------------
Consider {\it travelling solutions}
$(\psi_{\bf v},\chi_{\bf v})$
 with velocity ${\bf v}\in\R^3$, $|{\bf v}|<1$:
\beqn\label{9.8}
\left\{\ba{rcl}
\psi_{\bf v}(t,x)&=&S_{\bf v}\psi_0(\Lambda^{-1}_{\bf v}(t,x)),\\
\chi_{\bf v}(t,x)&=& \chi_0(y)
\quad \mbox{with }\,
y=x+{\bf v}\frac{(\gamma-1)}{|{\bf v}|^2} x\cdot{\bf v}-\gamma{\bf v}t.
\ea\right.
\eeqn
It is easy to check that $(\psi_{\bf v},\chi_{\bf v})$
is a solution of (\ref{8.1})--(\ref{8.2}).

Denote by ${\cal E}_{\bf v}:={\cal E}(\psi_{\bf v},\chi_{\bf v})$
the energy of the moving solitary waves $(\psi_{\bf v},\chi_{\bf v})$,
$$
{\cal E}_{\bf v}=
\int\left(\psi^*_{\bf v}(-i\al\cdot\nabla+m\beta)\psi_{\bf v}-
G(\bar\psi_{\bf v}\psi_{\bf v})+\frac1{2}
\Big(|\dot \chi_{\bf v}|^2+|\nabla\chi_{\bf v}|^2+M^2|\chi_{\bf v}|^2\Big)
-\eta\chi_{\bf v}\bar\psi_{\bf v}\psi_{\bf v}\right)\,dx.
$$
Assume that conditions {\bf C1} and {\bf C2} hold (see Section \ref{Sec3}).
Moreover, we impose the additional condition {\bf C3}.\\
{\bf C3}\, For given ${\bf v}=(v_1,v_2,v_3)\in\R^3$, $|{\bf v}|<1$,
\be\label{cond3}
\sum\limits_{i,j:\,i\not=j}\,v_iv_j\int\pa_i\chi_0(x)\pa_j\chi_0(x)\,dx\,=0.
\ee
The integral in (\ref{cond3}) equals
$(2\pi)^{-3}\ds\int\frac{k_ik_j|\hat f_\varphi(k)|^2}{(k^2+M^2)^2}\,dk$.
Then
condition {\bf C3} holds, for instance, if $\bar\varphi\varphi(x)$
is an even function in $x\in\R^3$.
In particular, conditions {\bf C1}--{\bf C3} are fulfilled  for the
particular family of solutions considered in Section \ref{Sec6.1.1}
(see also Section \ref{rem2.8} and formulas (\ref{2.13}) and (\ref{2.14})).

%%------------------------------------------------------------
\begin{lemma}
Let conditions {\bf C1}--{\bf C3} hold, ${\bf v}\in\R^3$ with $|\bf v|<1$.  Then
${\cal E}_{\bf v}=\gamma {\cal E}_0$.
\end{lemma}
{\bf Proof}
At first, consider the term in ${\cal E}_{\bf v}$ corresponding to the Dirac field
(cf formula (\ref{5.27})),
\beqn
{\bf e}_D&:=&\int\Big(\psi_{\bf v}^*(-i\al\cdot\nabla+m\beta)
\psi_{\bf v}-G(\bar\psi_{\bf v}\psi_{\bf v})\Big)\,dx
\nonumber\\
&=&\omega Q \gamma{\bf v}^2 +\frac1{\gamma}(I_1+I_2+I_3)+\frac1{\gamma}V
+\gamma\sum\limits_{j=1}^3v_j^2I_j+\eta_{\bf v},\nonumber
\eeqn
where $\eta_{\bf v}$ is defined in (\ref{etav}).
 %$:=-i\gamma \sum\limits_{j,k:j\not= k}(\varphi^*, \al_k \pa_j\varphi)v_k  v_j-2i\gamma(\varphi^*, \nabla\varphi)\cdot{\bf v}$.
Applying formula (\ref{IjKG}) and then the identity (\ref{8.3}), we obtain
\be\label{9.17}
{\bf e}_D=\gamma(I_1+I_2+I_3+ V)-\frac12\gamma{\bf v}^2R
-\gamma\sum\limits_{j=1}^3v_j^2P_j+\eta_{\bf v}.
\ee
Second, we rewrite the term in ${\cal E}_{\bf v}$ corresponding to the Klein--Gordon field,
\beqn\label{9.18}
{\bf e}_{KG}&:=&\frac1{2}\int
(|\dot \chi_{\bf v}(t,x)|^2+|\nabla\chi_{\bf v}(t,x)|^2+M^2|\chi_{\bf v}(t,x)|^2)\,dx\nonumber\\
&=&\frac1{2\gamma}\int
(|\nabla\chi_{0}(y)|^2+2\gamma^2|{\bf v}\cdot\nabla\chi_{0}(y)|^2+M^2|\chi_0(y)|^2)\,dy\nonumber\\
&=&\frac1{2\gamma}\int\chi_0(y)(-\Delta+M^2)\chi_0(y)\,dy
+\gamma\sum\limits_{j=1}^3v_j^2\int|\partial_j\chi_{0}(y)|^2\,dy+\eta'_{\bf v}
\nonumber\\
&=&\frac1{2\gamma}R+\gamma\sum\limits_{j=1}^3v_j^2 P_j+\eta'_{\bf v},
\eeqn
where, by definition,
$\ds\eta'_{\bf v}:=\gamma\sum\limits_{i,j:\,\,i\not= j}
v_iv_j\int\pa_i\chi_0(y)\pa_j\chi_0(y)\,dy$.
Further, the term in ${\cal E}_{\bf v}$ corresponding to the interaction is
\be\label{9.19}
{\bf e}_I:=-\eta\int\chi_{\bf v}(t,x)\bar\psi_{\bf v}(t,x)\psi_{\bf v}(t,x)\,dx
=-\frac\eta\gamma\int\chi_0(y)
\bar\varphi(y)\varphi(y)\,dy=-\frac1\gamma R(\varphi).
\ee
Applying (\ref{9.17})--(\ref{9.19}), and (\ref{9.14}), we obtain
$$
{\cal E}_{\bf v}={\bf e}_D+{\bf e}_{KG}+{\bf e}_I
=\gamma(I_1+I_2+I_3 +V)-\frac12\gamma R+\eta_{\bf v}+\eta'_{\bf v}
=\gamma {\cal E}_0+\eta_{\bf v}+\eta'_{\bf v}.
$$
Finally, by conditions {\bf C1}--{\bf C3}, $\eta_{\bf v}=\eta'_{\bf v}=0$.
Therefore, ${\cal E}_{\bf v}=\gamma {\cal E}_0$.\bo
%%---------------------------------------------
\medskip\\
{\bf Remark}\, If ${\bf v}=(0,0,v)$ with $|v|<1$, then
$\eta_{\bf v}=-2i\gamma v(\varphi^*,\partial_3\varphi)$ and $\eta'_{\bf v}=0$.
In this case,  conditions {\bf C2} and {\bf C3} are fulfilled, and
 condition {\bf C1} is equivalent to the condition
$(\varphi^*,\partial_3\varphi)=0$.
\medskip
%%--------------------------------

Denote by $P=(P^1,P^2,P^3)$ the momentum operator,
where $P^\beta=\int T^{0\beta}\,dx$,
 and $T^{\al\beta}$ ($\al,\beta=0,1,2,3,4$) denotes
 the energy--momentum tensor for the (KGD) model.
Using formula (13) from \cite{RanV} for the tensor $T^{\al\beta}$, we have
$$
P\equiv P(\psi,\chi)=-\int\Big(i\psi^*(t,x)\nabla\psi(t,x)
+\dot\chi(t,x)\nabla\chi(t,x)\Big)\,dx.
$$
 Put $P_{\bf v}:=P(\psi_{\bf v},\chi_{\bf v})$, ${\bf v}\in\R^3$.
 We impose conditions {\bf C1'} and {\bf C2}' on $\varphi_\omega$
 (see Section \ref{Sec3}).
  Moreover, we assume the additional condition {\bf C3'} on $\chi_0$.\\
  {\bf C3'}\,
  Let ${\bf v}=(v_1,v_2,v_3)\in\R^3$. For any $k=1,2,3$,
 $
\ds \sum\limits_{j:\,j\not=k}v_j\int\pa_k\chi_0(y)\pa_j\chi_0(y)\,dy
 %= (2\pi)^{-3}\int\frac{k_ik_j|\hat f_{\varphi}(k)|^2}{(k^2+M^2)^2}\,dk
 =0 $.

 Note that conditions {\bf C1'}--{\bf C3'} are fulfilled for the particular ansatz
 of solutions $\varphi\equiv\varphi_\omega$ considered in Section \ref{Sec6.1.1}.
% since for such solutions
% we have $\hat f_\varphi(k)=\eta F_{x\to k}[v^2(|x|)-u^2(|x|)]$
% and equalities (\ref{2.13}) and (\ref{2.14})  hold.

%%----------------------------------
 \begin{lemma} Let conditions {\bf C1'}--{\bf C3'} hold. Then
 $P_{\bf v}=\gamma{\bf v}{\cal E}_{0}$.
 \end{lemma}
%%%------------------------------------------------------
{\bf Proof}\, Using (\ref{9.8}) and (\ref{4.17}), we rewrite the term
in $P_{\bf v}$ corresponding to $\psi_{\bf v}$,
\beqn
P(\psi_{\bf v})&:=&-i
\int \psi^*_{\bf v}(t,x)\nabla\psi_{\bf v}(t,x)\,dx\nonumber\\
&=&-i\int\varphi^*(y)(\al\cdot{\bf v}+I)\Big(i\omega\gamma{\bf v}\varphi(y)+
\nabla\varphi(y)+{\bf v}\kappa\nabla\varphi(y)\cdot{\bf v}\Big)\,dy,\nonumber
\eeqn
where $\kappa=(\gamma-1)/(|{\bf v}|^2)$.
Applying (\ref{8.3}) and (\ref{IjKG}), we obtain (cf (\ref{5.36}))
\be\label{10.20}
P(\psi_{\bf v})={\bf v}\gamma{\cal E}_0-(P_1v_1,P_2v_2,P_3v_3)-
{\bf v}\kappa\sum\limits_{j=1}^3v_j^2P_j-i\xi_{\bf v},
\ee
where $\xi_{\bf v}$ is defined in (\ref{xiv}).
%\beqn
%\xi_{\bf v}&:=&{\bf v}\Big(\gamma+\frac{\gamma-1}{|\bf v|^2}\Big)
%\int\varphi^*\nabla\varphi\cdot{\bf v}\,dy+\int\varphi^*\nabla\varphi\,dy
%+{\bf v}\frac{\gamma-1}{|\bf v|^2}
%\sum\limits_{k\not=j}v_k v_j\int\varphi^*\al_k\pa_j\varphi\,dy \nonumber\\
%&&+\left(\sum\limits_{k\not=1}v_k\int\varphi^*\al_k\pa_1\varphi\,dy,\,
%\sum\limits_{k\not=2}v_k\int\varphi^*\al_k\pa_2\varphi\,dy,\,
%\sum\limits_{k\not=3}v_k\int\varphi^*\al_k\pa_3\varphi\,dy\right).\nonumber
%\eeqn
By conditions {\bf C1'} and {\bf C2}', $\xi_{\bf v}=0$.
%%-----------------------------------------------------------

Further, the second term
in $P_{\bf v}$ corresponding to $\chi_{\bf v}$ is
\beqn\label{10.21}
P(\chi_{\bf v})&:=&-
\int \dot\chi_{\bf v}(t,x)\nabla\chi_{\bf v}(t,x)\,dx
=\int(\nabla\chi_0(y)\cdot{\bf v})\Big(\nabla\chi_0(y)+{\bf v}\kappa
\nabla\chi_0(y)\cdot{\bf v}\Big)\,dy\nonumber\\
&=&(P_1v_1,P_2v_2,P_3v_3)+
{\bf v}\kappa\sum\limits_{j=1}^3v_j^2P_j+\xi'_{\bf v},
\eeqn
where $\xi'_{\bf v}$ stands for the following vector
\beqn
\xi'_{\bf v}&:=&
\left(\sum\limits_{j\not=1}v_j\int\pa_j\chi_0\pa_1\chi_0\,dy,\,
\sum\limits_{j\not=2}v_j\int\pa_j\chi_0\pa_2\chi_0\,dy,\,
\sum\limits_{j\not=3} v_j\int\pa_j\chi_0\pa_3\chi_0\,dy\right)\nonumber\\
&&+{\bf v}\kappa\sum\limits_{k,j:\,k\not=j}v_k v_j\int\pa_k\chi_0\pa_j\chi_0\,dy\, .\nonumber
\eeqn
By condition {\bf C3}', $\xi'_{\bf v}=0$. Hence, formulas (\ref{10.20}) and (\ref{10.21}) yield
$$
P_{\bf v}=-\int
\Big(i \psi^*_{\bf v}(t,x)\nabla\psi_{\bf v}(t,x)+\dot\chi_{\bf v}(t,x)\nabla\chi_{\bf v}(t,x)\Big)\,dx
=P(\psi_{\bf v})+P(\chi_{\bf v})=\gamma{\bf v}{\cal E}_{0}.\quad\bo
$$

%%%%%%%%%%%%%%%%%%%%%%%%%%%%%%%%%%%%%%%%%%%%%%%%%%%%%%%%%

\end{document}